\documentclass{aa}
\usepackage[varg]{txfonts}
\usepackage{graphicx}
\usepackage{newtxtext,newtxmath}
\usepackage{subcaption}
\usepackage{comment}
\usepackage[dvipsnames]{xcolor}
\usepackage{float}
\RequirePackage[colorlinks=true,linkcolor=blue,citecolor=blue,urlcolor=blue]{hyperref}

\newcommand{\mvir}{{\rm M}_{\rm 500c}}
\newcommand{\rvir}{r_{\rm 500c}}
\newcommand{\mtc}{{\rm M}_{\rm 200c}}
\newcommand{\rtc}{r_{\rm 200c}}
\newcommand{\msun}{\rm{M}_\odot}

\begin{document}

\title{What Drives Cluster Cool-Core Transformations?\\A Population Level Analysis of TNG-Cluster}

\titlerunning{Cool-Core Transformations in TNG-Cluster}

\author{Katrin Lehle\inst{1}\thanks{E-mail: k.lehle@stud.uni-heidelberg.de}
\and Dylan Nelson\inst{1}
\and Annalisa Pillepich\inst{2}
} 

\institute{Universit\"{a}t Heidelberg, Zentrum f\"{u}r Astronomie, ITA, Albert-Ueberle-Str. 2, 69120 Heidelberg, Germany \label{1}
\and Max-Planck-Institut f\"{u}r Astronomie, K\"{o}nigstuhl 17, 69117 Heidelberg, Germany \label{2}
}

\date{}

\abstract{In this study, we examine the frequency and physical drivers of transformations from cool-core (CC) into non-cool-core (NCC) clusters, and vice versa, in a sample of 352 massive galaxy clusters ($\mvir = 10^{14-15.3}~\msun$) from the TNG-Cluster magnetohydrodynamical cosmological simulation of galaxies. By identifying transformations based on the evolution of central entropy and focusing on $z\lesssim2.5$, we find that clusters frequently undergo such events, depending on their diverse assembly and supermassive black hole (SMBH) histories. On average, clusters experience two to three transformations. Transformations can occur in both directions and can be temporary, but those to higher entropy cores, i.e. in the qualitative direction from CC to NCC states, are the overwhelming majority. CC phases are also shorter than NCC phases, and thus overall the TNG-Cluster population forms with low-entropy cores and moves towards NCC states as cosmic time progresses. We study the role that mergers play in driving transformations, and find that mergers within $\sim 1$\,Gyr prior to a transformation towards higher (but not lower) entropy cores occur statistically more often than in a random control sample. Most importantly, we find examples of mergers associated with cool-cores disruption regardless of their mass ratio or angular momentum. However, past merger activity is not a good predictor for $z=0$ CC status, at least based on core entropy, even though clusters undergoing more major and minor mergers eventually have the highest core entropy values at $z=0$. We therefore consider the interplay between AGN feedback and evolving cluster core thermodynamics. We find that core transformations are accompanied by an increase in AGN activity, whereby frequent and repeated (kinetic) energy injections from the central SMBHs can produce a collective, long-term impact on central entropy, ultimately heating cluster cores. Whether such fast-paced periods of AGN activity are triggered by mergers is plausible, but not necessary.
}


\maketitle


\section{Introduction}
\label{sec_intro}
Galaxy clusters are the ultimate manifestation of the hierarchical growth of structures, forming from the subsequent merging of smaller objects over cosmic time and hence becoming the most massive gravitationally bound structures in the Universe. While dark matter constitutes most of their mass, the baryonic content is non negligible, as it asymptotes to the cosmic baryon fraction for the
most massive clusters. The baryonic mass component is dominated by the intracluster medium (ICM), a vast reservoir of hot, ionized gas that reaches extreme temperatures of $10^7-10^8\,$K \citep{mushotzky1978, mohr1999}.
The ICM is tremendously bright in X-ray, chiefly emitting via free-free bremsstrahlung \citep{lea1973}, allowing the cluster to loose its thermal energy. X-ray observations of the ICM reveal that some clusters have relatively low core entropy and short central cooling time compared to the Hubble time \cite{cowie1977, fabian1977}. This implies that the ICM in those clusters could rapidly cool down, condensate and stream into the core, forming a so-called `cool core'.

In contrast to these cool-core (CC) clusters, systems without such cool cores have also been observed, therefore labeled as non-cool-core (NCC) clusters \citep{molendi2001}. CC clusters are characterized by a prominent surface brightness peak, are at times associated with central drops in the temperature, and are often more relaxed \citep{vikhlinin2006, bartalucci2023}. In contrast, NCCs typically lack these features. Observations find both, CCs and NCCs, (similarly) frequently in their samples \citep{andrade-santos2017, hudson2007, graham2023}, although the exact fractions are under debate, as they strongly depend on the sample selection and the defining criterion of cooling status \citep{hudson2010, andrade-santos2017, eckert2011, lin2015}. 

The origin of the remarkable contrast between CCs and NCCs remains under debate, along with the questions whether they represent distinct sub populations and hence whether clusters undergo frequent transformations in core status over their lifetimes and, if so, what processes drive these transformations. A possible picture emerging from observations views CCs and NCCs not as distinct classes of clusters but as two different states in individual cluster histories that therefore can change. For example, \cite{rossetti2010} infer that most of the NCCs in their sample had a CC phase in their pasts. \cite{molendi2022} argues that, since the fraction of CCs remains roughly constant over time \citep[][but see selection caveats above]{ruppin2021, mcdonald2017}, the transformations of CCs into NCCs is balanced by a similar number of reverse transitions. In this framework, NCCs can evolve back into a CC cluster by cooling processes. However, the exact process is still unclear. It is also debated what processes destroy CCs. The prevailing theory suggests that, as NCC clusters are often more unrelaxed systems, CCs are disrupted by merging events \citep{allen2001, sanderson2006}. Since there exist NCC clusters that are relaxed, \cite{mccarthy2004} argues that, at least for those clusters, one has to consider other forms of non-gravitational entropy injection to explain the disruption of the cool core. For example, observations of several clusters have revealed strong AGN outbursts with energies ranging from $10^{61}$ to $10^{62}$ erg \citep{nulsen2005, mcnamara2009}, making it plausible that a cool core can be destroyed by heating of AGNs. In specific cases, such as in the galaxy cluster 3C 129.1, evidence suggests that a radio AGN outburst may have disrupted a previous cool core \citep{liu2024}.

Numerical simulations are crucial for unraveling the origins of cool and non-cool cores, as they allow us to easily trace the evolutionary history of galaxy clusters. Consistently with the observational considerations mentioned above, the most widely accepted scenario from simulations is that CCs and NCCs are simply different evolutionary phases, with clusters able to transition between the two states over time \citep[e.g.][]{poole2008, guo2009, barai2016}. Indeed, some simulations suggest that that CCs can transform into NCCs \citep{burns2007, poole2008, hahn2017, chadayammuri2021, villalba2024}. A potential physical mechanism to trigger such transitions are, also from the perspective of numerical models, mergers. Both cosmological and idealized simulations also show that not only the mass ratio but also the angular momentum (AM) plays a key role in determining whether a merger can destroy a cool core, with only low AM mergers \citep{hahn2017} or direct collisions \citep{poole2008} being capable of doing so. However, in idealized simulations of \citet{valdarnini2021} and \citet{chen2024} a more complicated picture arises and the survival of a cool core depends both on the initial mass ratio and AM. \cite{barnes2018}, by analyzing a relatively large population of clusters simulated in the full cosmological context and with relevant galaxy astrophysics processes, conclude that the fraction of relaxed clusters is similar for the CC and NCC populations, implying that mergers alone cannot be responsible for the disruption of some CCs. 

In addition to mergers, feedback from a central AGN can also inject substantial energy into the cores of clusters. At the same time, numerical simulations that do not include AGN feedback find that CCs are remarkably robust against mergers \citep{burns2007, poole2008}. This highlights the importance of AGN feedback in building realistic cool-cores through outflows that can soften the core i.e. increase central entropy. Such a process may boost the efficiency of core disruption through mergers or other channels. In idealized simulations, the ability of an AGN to destroy a cool core depends on the details of the physical model. For example, \cite{guo2009}, \cite{guo2010} and \cite{barai2016} find that AGNs can transform CCs into NCCs, while \cite{ehlert2023} report that AGN feedback from light jets cannot. \cite{chen2024} studied the combined effects of mergers and AGN feedback in idealized simulations of binary cluster mergers, finding scenarios where AGN feedback and mergers work together to destroy the core.

Simulations have shed some light on the physical drivers of cool-core transformations. However, this process has typically been studied in idealized setups, often without the background of a comprehensive galaxy formation physics model, or with cosmological simulations of small samples, typically $\mathcal{O}(1-10)$ clusters. Moreover, much focus in previous studies have been placed on the transformation of CCs into NCCs.

In this work we advance previous numerical studies, quantify the frequency and nature of cluster core transformations and explore their physical drivers. Our analysis employs the new cosmological magnetohydrodynamical simulation TNG-Cluster \citep{nelson2024}. This provides a large sample of hundreds of high mass galaxy clusters with cosmologically realistic merger histories. Importantly, these clusters have been shown to have broadly realistic ICM properties, from total gas content and X-ray luminosities \citep{nelson2024}, to CC statistics \citep{lehle2024} and size of the cooling radii (\textcolor{blue}{Prunier et al, submitted}). The AGN feedback model in TNG-Cluster also naturally produces a diversity of X-ray cavities comparable to observed cluster bubbles and cavities \citep{prunier2025}, including quantitative statistics of cavity properties (\textcolor{blue}{Prunier et al. submitted}), as well as turbulent velocities of Perseus-like clusters in agreement with Hitomi \citep{truong2024} and new XRISM observations (\textcolor{blue}{Truong et al. in prep}).

This study has three main objectives: (i) to demonstrate that clusters frequently undergo core transformations and can exhibit diverse thermodynamical histories; (ii) to highlight the role of merging events in driving core transformations; and (iii) to quantify the correlation between AGN feedback and core transformations. All this is done within a framework whereby AGN feedback is included and of the essence: it is invoked to quench star formation in massive galaxies \citep{weinberger2017, nelson2018, zinger2020}, including those at the center of galaxy clusters \citep{nelson2024}, and plays a key role in regulating the thermodynamics of the ICM at the center of clusters, i.e. at the center of their brightest cluster galaxy. 

This paper is organized as follows. Section~\ref{sec_TNGCluster} describes the TNG-Cluster simulation, Section~\ref{sec_def} summarizes all important quantities necessary for the analysis, while Section~\ref{sec_findTrafos} introduces our method to identify clusters core transformations. We present the merger identifying procedure in Section~\ref{sec_defMerger}. In Section~\ref{sec_stats} we present the results from applying the transformation selection process to our sample of 352 galaxy clusters of TNG-Cluster. In Section~\ref{sec_mergers}, we investigate the ability of mergers of driving transformation of cluster core. Subsequently, in Section~\ref{sec_agn}, we study the effects on AGN feedback on cluster cores. In Section~\ref{sec_discussion} we discuss the implications of our work and in Section~\ref{sec_conclusion} we summarize our findings.


\section{Methods} \label{sec_methods}

\subsection{The TNG-Cluster simulation} \label{sec_TNGCluster}

In this work we use the TNG-Cluster simulation \citep{nelson2024}, made up of $352$ high-resolution zoom simulations of high-mass galaxy clusters.\footnote{\url{www.tng-project.org/cluster}} It is a followup of the original three IllustrisTNG simulations, namely TNG50, TNG100, and TNG300 \citep[hereafter TNG;][]{nelson2018, pillepich2018a, marinacci2018, springel2018, naiman2018, pillepich2019, nelson2019b}, a suite of cosmological magnetohydrodynamical simulations designed to study galaxy formation and evolution. 

All of the TNG simulations including TNG-Cluster use the AREPO moving-mesh code \citep{springel2010} in order to solve the equations of ideal magnetohydrodynamics (MHD) \citep{pakmor2013, pakmor2011} plus self-gravity within an expanding spacetime. TNG-Cluster incorporates the TNG galaxy formation model and its included physics, unchanged. This is a well-validated model for much of the physics thought to be most relevant for galaxy evolution \citep[described in][]{weinberger2017,pillepich2018}. It includes gas processes such as heating from the metagalactic background radiation field (UVB), and radiative cooling from primordial plus metal species; star formation in dense gas; stellar evolution, chemical enrichment, and the return of mass and metals to the interstellar medium; stellar feedback, in the form of a galactic-scale wind model; and the formation i.e. seeding, mergers, and energetic feedback in thermal, kinetic, and radiative channels from supermassive black holes (SMBHs) i.e. active galactic nuclei (AGN). TNG-Cluster also adopts the same TNG cosmology: $\Omega_m = 0.3089$, $\Omega_b = 0.0486$, $\Omega_\Lambda = 0.6911$, $H_0 = 100 h$\,km\,s$^{-1}$\,Mpc$^{-1}$ = 67.74\,km\,s$^{-1}$\,Mpc$^{-1}$, $\sigma_8 = 0.8159$, and $n_s = 0.9667$ \citep[consistent with results of the][]{planckcollaboration2016}.

TNG-Cluster extends the scope and applicability of the TNG project by improving the available statistics i.e. sampling of the dark matter halo mass function at the high-mass end. The resulting cluster sample is (1\,Gpc) volume-complete at $M_{\rm 200c} \geq 10^{15}\msun$, and otherwise compensates for the rapid drop-off in statistics in TNG300 for $M_{\rm 200c} > 10^{14.5}\msun$ \citep[see][for details]{nelson2024}. The TNG-Cluster simulation has the same resolution as TNG300-1, with m$_{\rm{gas}} = 1.2\times10^7\msun$, m$_{\rm{DM}} = 6.1 \times 10^7\msun$, and a spatial resolution of $\sim 1$\,kpc. It was introduced in a series of papers presenting first science results and validations \citep{ayromlou2024,truong2024,lehle2024,lee2024,rohr2024,nelson2024}.

In the simulation, halos are identified using the standard friends-of-friends (FoF) algorithm with a linking length of $b=0.2$. We typically quote halo masses in terms of $\mvir$\footnote{$\mtc$ is the mass within $\rtc$, the radius of a sphere with an average density 200 times the critical density of the universe, at a given redshift. Similarly, $\rvir$ is the radius of the sphere with average density 500 times the critical density, and $\mvir$ is the corresponding mass enclosed.}. Substructures within FoF haloes are identified using the \textsc{SUBFIND} algorithm \citep{springel2001}, and we use the \textsc{SubLink} merger trees \citep{rodriguez-gomez2015} to link them across time.

\subsection{Definition of cluster properties} \label{sec_def}

\begin{figure*}
    \centering
	\includegraphics[width=0.95\textwidth]{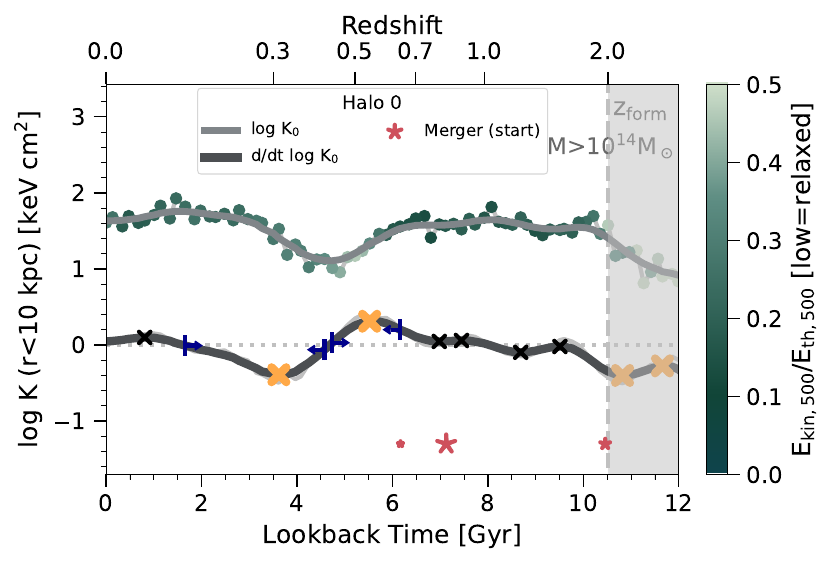}
    \caption{Representative time evolution of the central cooling state of a massive galaxy cluster. In particular, the central entropy of the most massive $z=0$ halo in TNG-Cluster is given by the circular symbols (color indicating relaxedness, in terms of the ratio of kinetic to thermal energy within $\rvir$). Correspondingly, the light gray line shows the smoothed time evolution of $K_0$, while the dark gray curve shows its time derivative. Local extrema of the derivative are indicated by crosses, with larger orange crosses marking significant core transformation events (see text). The blue markers indicate the timescales or duration of the two core transformations (see text for a definition). This halo undergoes a core transformation between $z \sim 0.3-0.5$ wherein the core evolves towards a more cool-core (CC) state. In this case the transformation is temporary and lasts $\sim 2-3$\,Gyr, after which the central entropy returns to its previous level. This type of behavior is one of several archetypes (see text for discussion). The pink stars mark the start of the three most massive merger events in the history of this cluster. The size of the markers indicate the mass ratio of the merging event. We mark the formation time of the cluster with the silver vertical, dashed line. Note that throughout the paper, time evolves from right to left in all plots that depict a time evolution.}
    \label{fig:K0vsTime}
\end{figure*}

Throughout this work we analyse the properties of the 352 galaxy clusters that are the main targets of TNG-Cluster. These cover a mass range of $\mvir = 1\times 10^{14}\msun$ to $\mvir = 1.9 \times 10^{15} \msun$, with an average mass of $\mvir = 4.3\times 10^{14}\msun$ at the current epoch ($z=0$). For each, we measure several physical quantities.

\paragraph{\textbf{Formation redshift}}

We modify the usual definition of formation redshift as the time when the halo reached half its current mass. Instead, we define the formation redshift of a cluster as the time when its main progenitor first reaches a mass of $\mvir=10^{14}\msun$. This is a canonical threshold for the minimum halo mass to be labeled a cluster, so our definition corresponds to the time when the progenitor has a cluster-like dark halo mass and thus potentially cluster-like physical properties in its ICM.

\paragraph{\textbf{Thermodynamic properties}}

Our analysis is built around the time evolution of the central cooling state of clusters. As our principal physical quantity for this purpose, we use central entropy. We compute the entropy of gas as $K = k_{\mathrm{B}}  T  n_{\mathrm{e}}^{-2/3}$. The central entropy $K_0$ is calculated from all gas within a 3D aperture $r<10$\,kpc, centered on the position of the gravitational potential minimum. We include only (i) non-star-forming gas, (ii) gas with net cooling, and (iii) gas with a minimum temperature $T>10^6$\,K. This roughly corresponds to the X-ray emitting gas from which observational estimates of $K_0$ are derived.

\paragraph{\textbf{Cooling status}}

Unless stated otherwise, to define the cooling state of a cluster we use the observationally-motivated thresholds of \cite{hudson2010}. Strong cool-core clusters (SCCs) have $K(r<10$\,kpc$) \equiv K_0 \leq 22$\,keV\,cm$^2$, weak cool-cores (WCCs) are clusters with 22\,keV\,cm$^2 < K_0 \leq 150$\,keV\,cm$^2$, and non-cool-cores (NCCs) have $K_0> 150$\,keV\,cm$^2$. 
Details regarding these definitions, and the level of (dis)agreement between different definitions of CC status in TNG-Cluster can be found in \cite{lehle2024}\footnote{Note that \cite{lehle2024} use a different aperture for the definition of the central entropy. Here, we have chosen a fixed physical aperture that is independent of the virial properties of the cluster. We made this choice to factor out the time dependence of $K_0$ on halo mass, when defined via a fraction of $\rvir$ (see Fig.~12 of that work).}.
\paragraph{\textbf{Relaxedness}}

We define the relaxedness of a cluster based on an energetics argument, by comparing the kinetic to thermal energy of the ICM \citep[following][]{barnes2017a}. In particular, a cluster is relaxed if $E_{\mathrm{kin, 500c}}/E_{\mathrm{th, 500c}} < 0.1$, 
where $E_{\mathrm{kin, 500c}}$ ($E_{\mathrm{th, 500c}}$) is the sum of the kinetic (thermal) energy of all gas within $\rvir$. 

\subsection{Identifying cluster transformations} \label{sec_findTrafos}

In this work we aim to identify significant changes in central entropy as signposts of cool-core state transformations. We develop an automatic classification algorithm for this purpose, which does not require visual inspection or manual decisions for individual halos. Furthermore, we aim for a classification independent of specific entropy thresholds, since the commonly used thresholds do not coincide with features in the distribution of central values \citep{lehle2024} and their validity at higher redshifts is unclear. Also, we found that strong and significant changes in the central properties of the clusters during their lifetime do not align with the thresholds used in the literature. Below, we outline the steps of our method, while Fig.~\ref{fig:K0vsTime} provides a visualization of this procedure\footnote{Time evolves from right to left in all plots of the paper.}. 

\begin{enumerate}[(i)]
    \item We start by considering the time evolution of the central entropy, $K_0(t)$. Entropy is selected because it is readily available in TNG-Cluster and observations and is well-suited to trace the cooling status of a cluster \citep{hudson2010, lehle2024}.
    \item To avoid spurious short time-scale features, we smooth $K_0(t)$ using a 1D Gaussian filter with a standard deviation of the Gaussian kernel of 2.5 and a filter size of 10 (light grey curve in Fig.~\ref{fig:K0vsTime}). This smoothing procedure effectively results in a smoothing window of $\lesssim 1\, $Gyr size. 
    \item Next, we compute the derivative of the smoothed $K_0(t)$ and, again, smooth the derivative with the same Gaussian filter (dark grey curve in Fig.~\ref{fig:K0vsTime}). 
    \item We then identify all local extrema in the smoothed derivative (marked by black crosses in Fig.~\ref{fig:K0vsTime}). These extrema indicate all points of change in the history of a cluster. 
    \item Since we are only interested in `significant' changes of the central entropy, we apply two selection criteria to the extrema. First, we consider changes only after the formation of a cluster (the formation redshift is marked by the dashed vertical silver line in Fig.~\ref{fig:K0vsTime}). Second, we select extrema only if they exceed a certain constant threshold in the derivative. For our primary analysis, we set this threshold at 0.19, a value\footnote{Given that the results of our analysis depends on this threshold and there is no a priori correct choice, we repeat our analysis throughout, varying both lower (0.13) and higher (0.25) thresholds, to assess the sensitivity of our results to this parameter.} calibrated to successfully include clear extrema. The selected times are marked by yellow crosses in Fig.~\ref{fig:K0vsTime}. 
\end{enumerate}

\subsubsection{Timescales of transformations}\label{sec_def_timescale}

For each transformation, we define the `transformation timescale' as the duration of the event. Specifically, we identify a starting time and ending time by searching on both sides of each transformation event (yellow cross) for one of two criteria:

\begin{enumerate}[(i)]
\item The time when the derivative changes sign. 
\item The midpoint between the adjacent local extrema (considering all, i.e. both black and yellow crosses in Fig.\ref{fig:K0vsTime}). 
\end{enumerate} 

For both the starting and ending times we adopt the closest such point. For our case study, the starting and ending times of the two transformations are marked with blue indicators in Fig.\ref{fig:K0vsTime}. They are $\sim 1$\,Gyr and $\sim 2.5$\,Gyr long, respectively. In this case, it is clear that the overall evolution of central entropy experiences a drop, followed by a rise back to roughly its pre-drop value. Our (two) transformations correspond to the drop and the rise, respectively. The transformation timescales correspond to the lengths of these two events, respectively, and not to the duration of the entire low-$K_0$ episode, which we analyze separately.

\begin{figure*}
    \centering
	\includegraphics[width=0.4\textwidth]{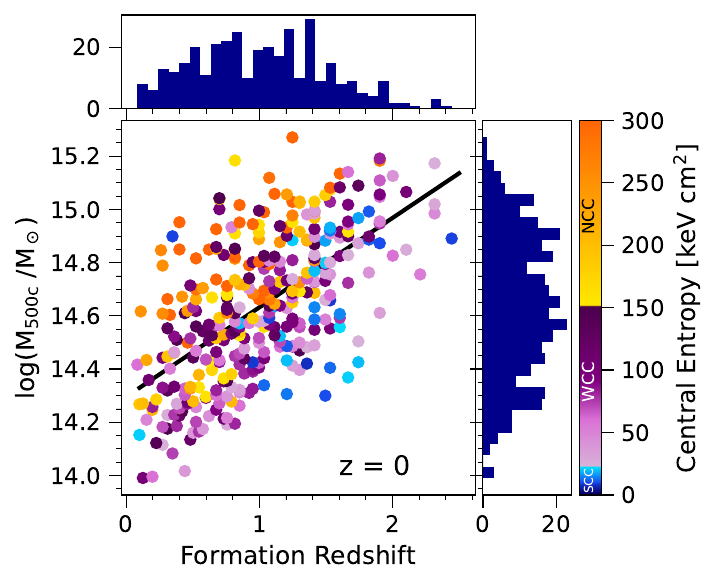}
    \includegraphics[width=0.4\textwidth]{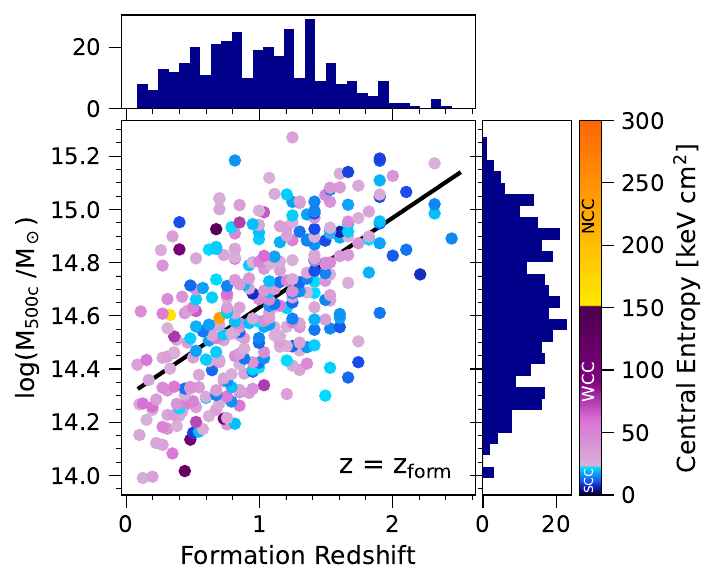}
    \includegraphics[width=0.45\textwidth]{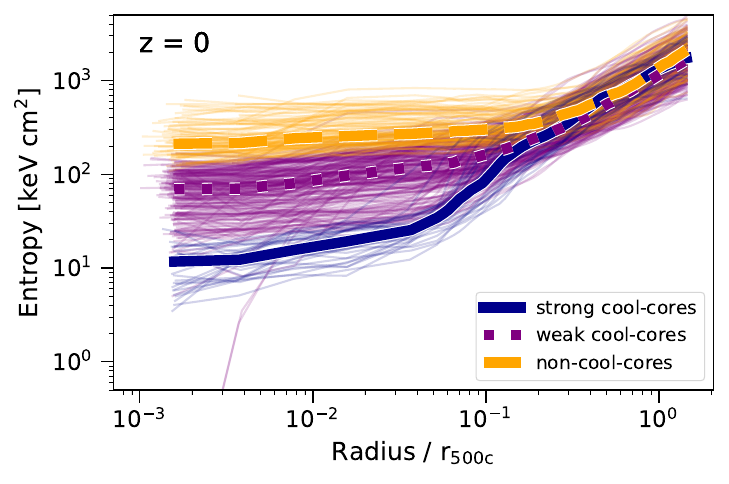}
    \includegraphics[width=0.45\textwidth]{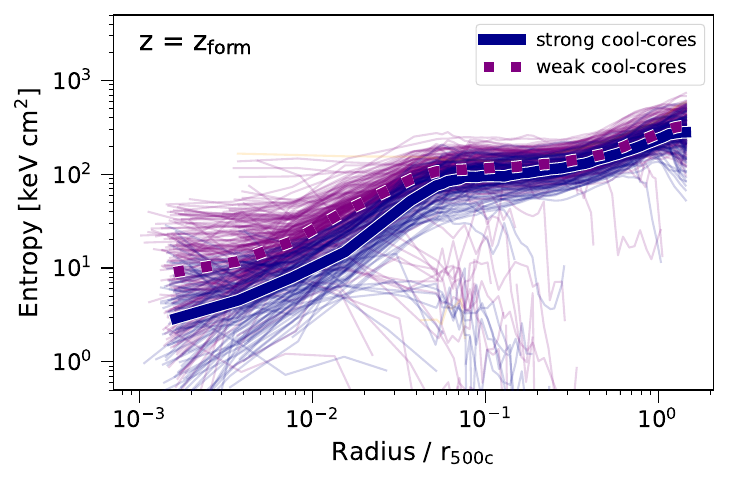}
    \caption{Correlation among halo mass $\mvir$ at $z=0$, formation redshift and cool-coreness in TNG-Cluster, across cosmic epochs. The formation redshift throughout this paper is defined as the redshift at which a cluster reaches a mass of $10^{14} \msun$. The upper panels show the correlation of formation redshift and the halo mass at $z=0$. Symbols are color-coded by the central entropy of each halo at $z=0$ (left plot) and at the formation redshift $z_{\rm form}$ (right panel), with the color bar indicating the classification into non-cool-core (NCC), weak cool-core (WCC), and strong cool-core (SCC) clusters. Histograms of the distribution of formation redshifts and cluster masses are shown on the top and right panels, respectively. The lower plots present entropy profiles at $z=0$ (left) and at $z_{\rm form}$ (right) colored by core status. More massive halos form, i.e. reach the status of cluster, on average earlier than less massive halos. For a given halo mass, later forming halos are biased towards NCC states at $z=0$. In addition, clusters form with low central entropy, with SCCs and WCCs having similarly shaped profiles exhibiting a prominent decline towards the core. According to TNG-Cluster, clusters are predominately born as CCs.}
    \label{fig:M500VsZform}
\end{figure*}

\begin{figure*}
    \centering
    \includegraphics[width=0.9\linewidth]{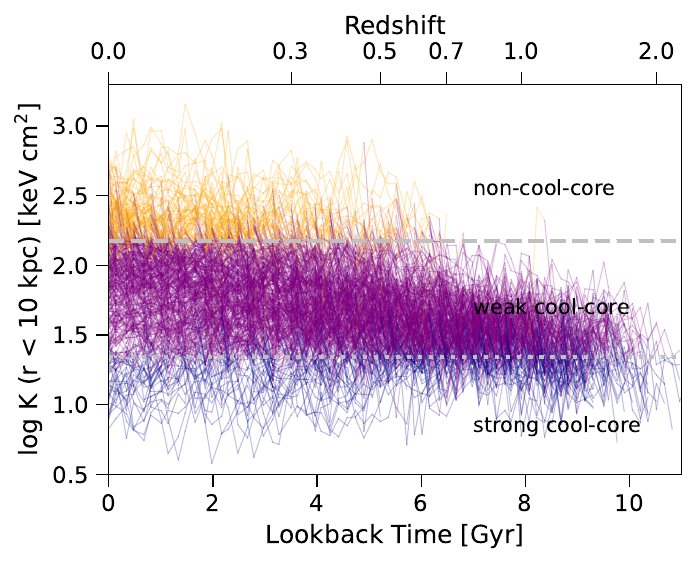}
    \caption{Time evolution of the central entropy for all halos of TNG-Cluster after the formation of each cluster. Each line represents one of the 352 clusters as it evolves and grows in mass through time. Lines are colored by the core status at each redshift. The overall cluster population is dominated by SCCs and WCCs at high redshift and gains progressively more NCC towards low redshift. Amid the overall evolution towards the NCC regime, there is a huge diversity in the central entropy evolution from cluster to cluster.}
    \label{fig:K0vsTimeAllHalos}
\end{figure*}

\subsection{Identifying merging events}\label{sec_defMerger}

We identify mergers between the progenitor halos of each cluster, for all 352 clusters throughout their lifetime (i.e. for $z < z_{\rm form}$). We define the time of merger events as the time when the secondary crosses $0.5 r_{\rm 500c}$ of the main cluster. To do so, we use the SubLink merger tree \citep{rodriguez-gomez2015}, and linearly interpolate between snapshot times. We use this crossing time to mark the beginning of the merging event, i.e. when the merger can plausibly start to have a significant impact on the thermodynamical state of the cluster core. 

Over the duration of a merger, as the secondary falls in to the potential well of the main cluster, it starts loosing mass. To avoid ambiguity, we therefore compute the merger mass ratio $\rm M_{\rm 500c, sub}/\rm M_{\rm 500c, main} (t_{\rm maxMass})$ at the time of the maximal mass of the secondary. Throughout this work, we only consider mergers with mass ratios of 0.01 or larger. In addition, we label major mergers as those with mass ratios above 0.25, while minor mergers have mass ratios above 0.1. Mini mergers include all mergers with mass ratio above 0.01 and below 0.1.

Finally, depending on angular momentum of the collision, the effects on the core of the cluster may differ. Thus, as a proxy of angular momentum we compute the tangential velocity of the secondary when it crosses $0.5r_{\rm 500c}$ of the main cluster. The tangential velocity $v_t$ is defined as
\begin{equation}
    v_t = |\vec{v}_{\rm tot} - (\vec{v}_{\rm tot} \cdot \vec{e}_r)\vec{e}_r |\, .
\end{equation}
Here, $\vec{v}_{\rm tot}$ is the total velocity vector and $\vec{e}_r$ is the radial unit vector.

\noindent We consider the tangential velocity relative to the circular velocity of the main cluster:  
\begin{equation}
    v_{\rm c} = \frac{G \mvir}{\rvir} \, .
\end{equation}

\noindent Small tangential velocities, that is low $v_t/v_c$, correspond to low angular momentum encounters i.e. more radial collisions.


\section{Statistics of cluster transformations} \label{sec_stats}

We begin with the population-level demographics of core transformations in TNG-Cluster. 

\subsection{Clusters form with low-entropy cores and collectively move towards NCC states at lower redshift}
We first give a general overview of the overall evolution of the total halo population in TNG-Cluster. Furthermore, as our study considers the evolution of cluster progenitors starting from their formation redshifts, we also explore how formation redshift correlates with halo properties, such as halo mass, in order to contextualize our subsequent analyses of cluster transformations.

The upper left panel in Fig.~\ref{fig:M500VsZform} shows the relationship between halo mass $\mvir$ and formation redshift at $z=0$, with each marker color-coded by the current core entropy. Clusters form on average at $z \simeq 1.0$, although individual halos span a broad range from $z=0.08 - 2.4$\footnote{Under the more usual definition of formation redshift, i.e. as the time when a halo reached half its current mass, the average formation redshift of our sample is instead $z=0.52$.}. As expected, i.e. given our mass threshold-based definition of formation time, more massive clusters formed earlier than less massive clusters. 

In addition, there are strong trends of $z=0$ CC state with both mass, at fixed formation redshift, and with formation redshift, at fixed mass.
In particular, NCC clusters (orange) occupy the upper left corner of the diagram, indicating that these clusters are either more massive or formed at later times. Conversely, SCCs (blue) are predominantly found in the lower right side of the parameter space  -- they are less massive and/or formed at high redshift. One interpretation suggested by previous literature results is that halos forming early have more time to relax, while less massive halos experience fewer disruptive mergers, allowing both to maintain a more relaxed, and hence SCC, state. WCCs (purple) occupy intermediate masses and formation redshifts. 

The upper right panel shows the same data, but now colored by the core status of the clusters at the time of their formation. This plot clearly shows that, according to TNG-Cluster, clusters are born with significantly lower core entropy than at the current epoch. At the time of their formation, 124 clusters are SCCs, 226 are WCCS, and only 2 clusters are NCCs. On the other hand, 106 of the 352 are NCCs at  at $z=0$. Moreover, most of the WCCs at the formation redshift have central entropy values very close to the threshold separating SCCs from WCCs. These exact numbers and figures certainly depend on the mass distribution of the TNG-Cluster sample (which is volume-limited only at the highest-mass end) and are based on the CC entropy thresholds that are more appropriate of the low-redshift Universe. The exact proportions may be different for differently-selected cluster samples and for different definitions of core status. Nevertheless, it is clear that overall the cluster population gains a larger and larger fraction of NCCs as the time progresses. 

The findings above are supported by the entropy profiles presented in the lower two panels of Fig.~\ref{fig:M500VsZform}. The left plot shows the entropy profiles at $z=0$, while the right one shows the profiles at the time of the formation of the clusters. Whereas, there is a diversity in entropy profiles at $z=0$ with SCCs and WCCs/NCCs having well separated and differently shaped profiles, the profiles at $z = z_{\rm form}$ mainly have similar shape, with a pronounced dip to low entropy in the core. 

Fig.~\ref{fig:K0vsTimeAllHalos} summarizes the time evolution of cluster cores by visualizing the evolution of the central entropy, as depicted in Fig.~\ref{fig:K0vsTime} for an individual halo, but across the whole cluster population in TNG-Cluster. Globally, the whole population evolves from a SCC- and WCC-dominated sample at high redshifts to a more diverse population towards the NCC regime. This is in agreement with \citet[][Fig. 12]{lehle2024}\footnote{\citet{lehle2024} employed a different aperture definition for measuring central entropy, which depends on the virial radius. This choice leads to a stronger redshift evolution of central entropy compared to the evolution observed with the aperture sizes used in this paper. We reproduced their Fig.~12 using our aperture definition and found similar, though slightly weaker, trends.}. 

Observations of galaxy clusters up to $z=2$ reveal little to no evolution in central entropy \citep{mcdonald2013, mcdonald2017, sanders2018, pascut2015}. For example, \citet{mcdonald2013} found little variation in $K_0$ with increasing redshift, but with their cluster sample maintaining a consistent average mass across the redshift range. \citet{lehle2024} reported that, when focusing exclusively on high-mass clusters, their trend of central entropy is consistent with no evolution. Building on this, we find that for halos with $M>10^{15}\msun$, some of the cluster exhibit little overall entropy evolution, while others transition from SCCs to NCCs. As new clusters join the sample at each redshift, by crossing the mass threshold, the overall $K_0$ distribution remains largely unchanged. Both effects combined lead to a high-mass, mass-limited cluster sample that does not evolve towards the NCC regime overall for the highest mass clusters. These considerations highlight the significant influence of sample selection on the redshift evolution of a cluster population.

The results of Figs.~~\ref{fig:M500VsZform} and \ref{fig:K0vsTimeAllHalos} are all consistent with a cluster population that collectively, and by growing in  cluster mass, moves towards NCC states as cosmic time progresses, with clusters been predominantly born as CCs. Note, however, that these three states and three distinct colors artificially separate the cluster population into three disjoint classes, while we instead find that the distribution of $K_0$ is unimodal with no clear motivation to separate clusters based on $z=0$ central entropy \citep[][Fig. 7]{lehle2024}. Given these circumstances, we base the analysis in the rest of this work on methods independent of the entropy thresholds of CCs vs. NCCs used in the literature.

\subsection{Frequency and timing of core transformations across the TNG-Cluster population}

We now apply our algorithm to identify cluster core transformations (see Sec.~\ref{sec_findTrafos}). 

In total we find 1874 changes in the core entropy derivatives (i.e. black and yellow crosses in Fig.~\ref{fig:K0vsTime}) of the 352 clusters of TNG-Cluster: of these, 1249 are negative (i.e. changes to higher core entropy, i.e. \textit{in the direction} from CC to NCC status) and 625 are positive (i.e. changes to lower core entropy, i.e. \textit{in the direction} from NCC to CC status). Applying our fiducial threshold for significance (see discussion in Appendix \ref{appendix_deriv}), we arrive to a final sample of 478 (26\%) negative and 97 (5\%) positive core transformations. Broadly speaking, clusters frequently undergo transformations in both directions, toward higher and lower core entropy, though transitions to higher core entropy are more common according to TNG-Cluster than the reverse.

\begin{figure}
    \centering
	\includegraphics[width=0.46\textwidth]{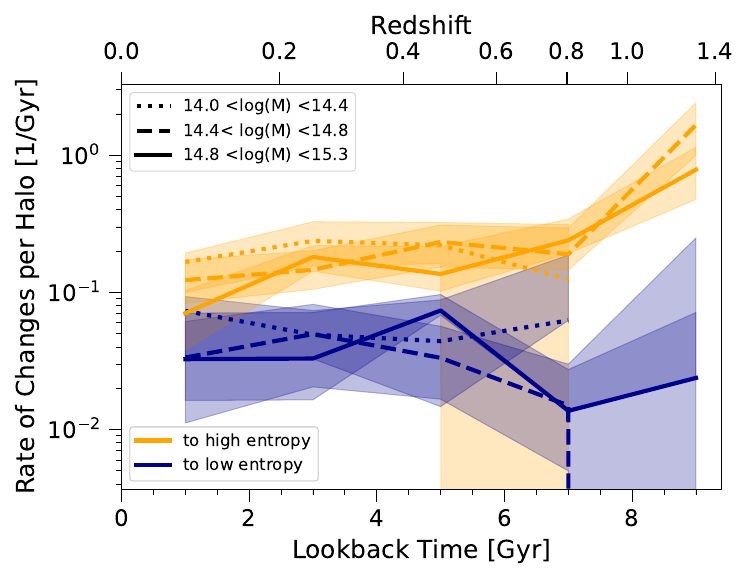}
    \caption{Rate of core changes as a function of lookback time for three different mass bins according to TNG-Cluster. We define the rate as the number of transformations in a mass and time bin divided by the length of the time interval and number of halos in that bin (details in the text). Orange (blue) lines indicate the rate of changes to higher (lower) core entropy. The bands for each line mark the methodological uncertainty, that is, the variation of the result we obtain when redoing the analysis with a larger and lower threshold of the entropy time derivative. The solid lines show the result for our fiducial choice of threshold. We find no trend with mass and only very weak redshift trends for both type of transformations, but the rate for changes to lower core entropy is overall smaller.}
    \label{fig:ChangeRate}
\end{figure}

Fig.~\ref{fig:ChangeRate} shows the rate at which transformations occur for halos of three different mass bins as a function of redshift. This rate is defined as the number of changes in a mass and time bin divided by the length of the time bin (here $2\,$Gyr) and the number of halos in that bin. The rates for both, changes to higher (orange) and lower (blue) core entropy, show no significant differences for the three mass bins and only mild redshift dependencies. The rate for changes to lower entropy is overall smaller, since we have overall fewer of those transformations. 

The rate for changes to higher core entropy mildly increases with redshift, from $\sim 0.1$ per billion year at recent times to $\sim 0.2$ Gyr$^{-1}$ at $z\sim 0.5$, and steeply increasing in the earliest time bin to about unity. The rate for transformations to lower core entropy is more noisy but consistent with no redshift evolution, with values of $\sim 0.03$ Gyr$^{-1}$. Different selection thresholds result in the same qualitative trends. 
As transformations towards high entropy occur more frequently, this implies that the global CC fraction of the cluster population should decrease with time i.e. towards $z=0$. This is in agreement with the findings highlighted above and with those by \citet[][their Figs. 12 and 13]{lehle2024}. In TNG-Cluster, they find that the SCC fraction decrease towards $z = 0$, with a strong decrease in low- and intermediate-mass bins but a rather constant trend for the most massive clusters.

\begin{figure}
    \centering
	\includegraphics[width=0.46\textwidth]{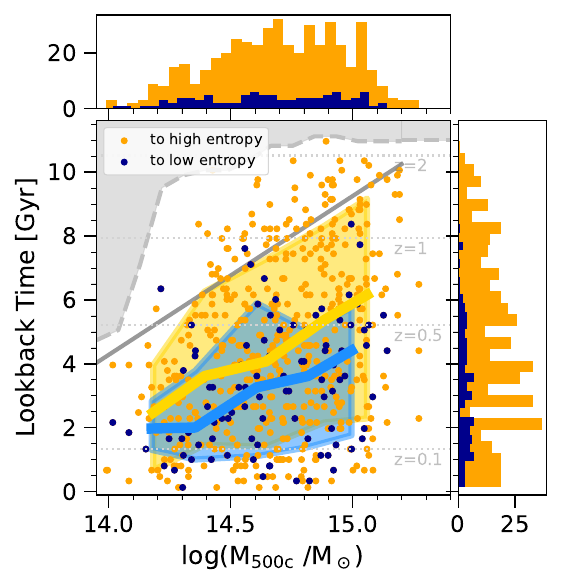}
     \caption{Correlation between the lookback time of a transformation and cluster mass at $z=0$ according to TNG-Cluster. Transformations to higher core entropy are shown in orange, while transformations to lower core entropy are shown in dark blue. Marginalized histograms of each axis quantity are shown with the top and right subpanels, respectively. The gray dashed line indicates the earliest formation times for our halos, i.e. the earliest possible transformation times. The solid grey line depicts the average formation times (i.e. the fit to the results presented in Fig.~\ref{fig:M500VsZform}). The gold (light blue) lines and bands show the median and 16 to 84 percentile ranges for changes to higher (lower) core entropy. The average lookback time of transformations increases with halo mass, in part because more massive clusters form earlier and may therefore experience more and earlier transformations. Changes to higher core entropy happen at all times, while changes to lower core entropy occur only at later times, potentially reflecting different physical drivers.}
    \label{fig:M500vsLbtTrafos}
\end{figure}

Finally, core transformations occur throughout the evolutionary lifetime of clusters. Fig.~\ref{fig:M500vsLbtTrafos} shows the relationship between transformation time and $z=0$ halo mass, with color indicating changes towards either higher (orange) or lower (blue) central entropy. The median trends for each subset are shown with thick colored lines, and in general we see that in both cases transformations typically occur earlier for more massive halos: i.e. the median lookback time increases as a function of $z=0$ halo mass. This result is expected as more massive clusters form earlier on average, given our definition, and thus may undergo earlier transformations. It may also reflect the relatively high frequency of merger events at earlier times, as we discuss further below.

The gray dashed line shows an upper envelope as a function of halo mass, indicating the earliest halo formation times and thus earliest possible transformations that we identify. The solid grey line indicates the average formation time as a function of halo mass, which is the fit to the values in Fig.~\ref{fig:M500VsZform}. At the same time, the right subpanel shows the marginalized distribution of transformation lookback times. We see that the distributions for both types of transformations are broad with a mean of lookback times of 4.6\,Gyr (3.3\,Gyr) and standard deviation of 2.7\,Gyr (1.9\,Gyr) for changes to higher (lower) core entropy. Notably, changes to high entropy occur throughout all of cosmic time, while changes towards low entropy happen preferentially near redshift zero.\footnote{If we consider only the first or last transformation that each cluster experiences, the results are qualitatively unchanged, although the mean timings shift earlier or later, respectively (not shown).}

\begin{figure}
    \centering
	\includegraphics[width=0.48\textwidth]{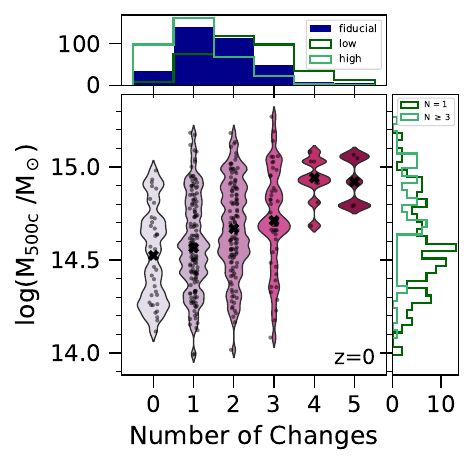}
     \caption{The number of core transformations across the history of each TNG-Cluster halo, as a function of $\mvir$ at $z=0$. In the main panel, black dots represent individual halos, while the equal-width violins depict the corresponding mass distributions. The black crosses show the median halo mass for each number of transformations. The subpanels show the distributions of number of transformations (top) and halo mass (right). In the top, we contrast our fiducial analysis choice (dark blue) with the impact of varying the threshold applied to the $K_0$ time derivative (green and light blue histograms). The histogram in the right panel shows the mass distribution for clusters that have only one (equal or more than three) transformation(s) in darker (lighter) green. In general, the majority of clusters undergo between one and three transformations over their lifetime. More massive clusters undergo more transformations, on average.} 
    \label{fig:M500VsNumTrafo}
\end{figure}

\begin{figure*}
    \centering
     \includegraphics[width=0.8\textwidth]{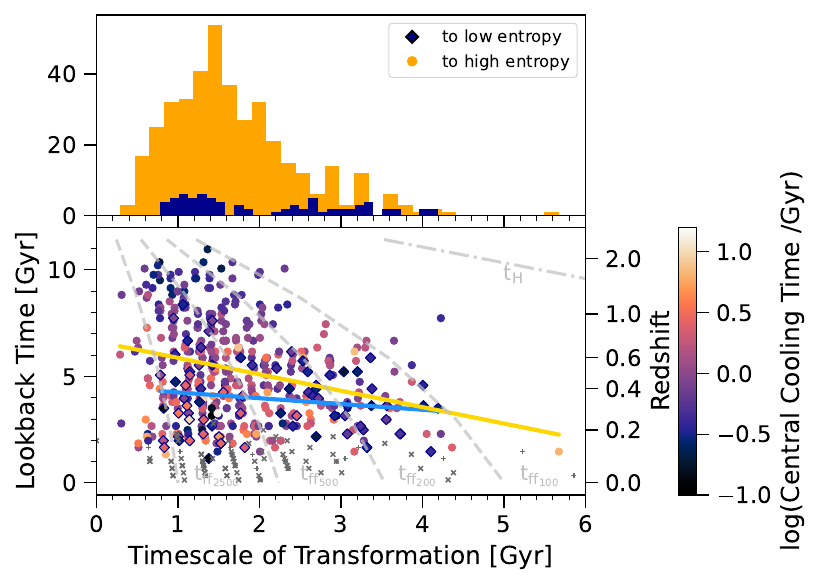}
     \caption{Correlation of transformation timescale with the lookback time of the transformation (lower panel) and the marginalized distribution of these timescales (upper panel) across the total TNG-Cluster population. The timescale measures the duration of a transformation (for a detailed definition see Sec.~\ref{sec_def_timescale}). In the top panel, the yellow histogram shows transformations towards high entropy, while blue shows transformations to low entropy. In the main panel, circles show the former, and diamonds with blue outlines the latter, while all symbols are colored by central cooling time at the time of the transformation. Small gray crosses and pluses mark transformations near $z=0$ for which we cannot measure a timescale (see text). The dash-dotted line indicates the age of the universe as a function of redshift, and the dashed lines show free-fall times at various radii. Transformations to higher core entropy are on average faster with a mean (standard deviation) of 1.7 Gyr (0.8 Gyr), versus changes to lower core entropy (mean: 2.0 Gyr, standard deviation: 1.0 Gyr). The distribution of transformation timescales towards low entropy appears double-peaked (see text).}
    \label{fig:TimescaleChange}
\end{figure*}

In general, we find that, at the time of the first transformation, clusters that change to lower core entropy have a relatively high central entropy. Further, the lookback time of the first transformation to higher core entropy correlates well with $K_0$ at that time. The larger the central entropy at the first transformation, the later the first transformation to higher central entropy happens (not shown). We interpret this primarily as a mass and cosmic time trend, as halo mass and also central entropy increase across the cluster population towards $z=0$. Therefore, transformations happening late are found at higher central entropy. As another result, the onset of transformations towards lower entropy occurs later, by several Gyr of lookback time, than the onset of transformations towards higher entropy. 

\subsection{Core transformations in individual cluster histories}
We have so far discussed all transformations across 352 TNG-Cluster halos, where an individual cluster can (and often does) undergo more than one such transformation during its lifetime. 

Fig.~\ref{fig:M500VsNumTrafo} shows the number of transformations \textit{per cluster} as a function of halo mass $\mvir$ at $z=0$. In general, more massive clusters undergo more core transformations. Only the most massive halos in TNG-Cluster have $\geq 4$ such events over their lifetime. The median halo masses for each number of transformations (black crosses in Fig.~\ref{fig:M500VsNumTrafo}) likewise increases from $\sim 10^{14.5}\msun$ for $N=0$, by 0.2\,dex for $N=3$, and by a further 0.2\,dex to $\sim 10^{14.9}\msun$ for $N \geq 4$. The subpanel on the right shows the marginalized distributions of the halo mass for clusters that have exactly one transformation (darker green) and have at least three transformations (lighter green). The former distribution peaks at $10^{14.58}\msun$, while the median of the latter one is at $10^{14.8}\msun$. While more massive clusters have had a longer time `as a cluster' to undergo transformations, they are also built from more complex merger histories, and host more massive SMBHs, that have therefore injected more cumulative feedback energy. We explore the role of both processes in driving core transformation below.

The upper panel gives the marginalized distributions of the number of transformations across the population. Adopting our fiducial definition (dark blue), the majority of clusters have one (141) or two (113) transformations. These numbers strongly depend on the threshold of the $K_0$ derivative chosen. Specifically, if we adopt our lower (higher) threshold, clusters have on average $\sim 2-3$ (or $\sim 0-1$) transformations.

\subsection{Timescales of cluster core transformations}

To help understanding the physical drivers of core transformations, we assess the timescales i.e. overall duration of such events, adopting the definition and method presented in Sec.~\ref{sec_def_timescale}. 

Fig.~\ref{fig:TimescaleChange} shows the distribution of transformation timescales (top panel), as well as the correlation with the lookback time when the transformation occurred (bottom panel). The distribution of timescales for changes to higher core entropy (orange) is single peaked with a tail towards longer times. The average duration is 1.7\,Gyr. In contrast, the distribution of transformations to lower core entropy (blue) is broader and appears to be double-peaked, with the two peaks separated at $\sim2$\,Gyr. A Gaussian Mixture Model suggests this distribution is better fit by a bimodal distribution than an unimodal one. 

In the main panel, each transformation is represented by a marker: circles denote transformations towards high entropy, and diamonds vice versa. 
Linear fits reveal that earlier transformations occur faster, with shorter timescales ($<2$\,Gyr) seen across all epochs, while slower transformations ($>2$\,Gyr) occur only at $z<1$. This trend is stronger for changes to higher core entropy. \footnote{At late times we cannot measure transformation timescales if they occur near $z=0$: these events are marked by dark gray crosses (pluses) for transformations to high (low) core entropy. For the gray markers we assume symmetry and compute the timescale as $2\times (t_{\rm start}-t_{\rm trafo})$. Those 102 transformations are not included in the upper histograms.}

To put the timescale of transformations into context, we also show the Hubble time as a function of redshift (dash-dotted line) and the average free-fall timescale at $r_{\rm 2500c}$, $\rvir$, $\rtc$ and $r_{\rm 100c}$ (dashed lines).\footnote{We derive the average free-fall timescale as a function of redshift as $\rm M_\Delta \left( \frac{4\pi}{3}\rm r_\Delta^3\right)^{-1} = \Delta \rho_{\mathrm{cr}}(z) = \Delta \frac{3H(z)^2}{8\pi G}$. In this case $t_{\mathrm{ff}_{\Delta}}(z) = \frac{12}{\Delta} t_H (z)$.} We find that transformation timescales are correlated with the free-fall time, although there is a large scatter. The free-fall timescales from $r_{\rm 2500c}$ to $r_{\rm 100c}$ bracket the range of timescales we find, suggesting that core transformations occur on timescales similar to the dynamical times at these scales. 

We also color the markers in the lower panel by the cooling time at the time of the transformation.\footnote{At snapshots where the cooling time is not directly available, we derive it from the correlation with central entropy at the nearest time. At tight correlation between central cooling time and central entropy is expected and has been shown in several works (e.g. \cite{hudson2010}, or \cite{barnes2018} and \cite{lehle2024} for the TNG model.} The cooling times at higher redshift are on average shorter than at lower redshift, as to be expected \citep[see e.g.][for a discussion]{rohr2024a}. For $z < 0.6$ and for changes to lower core entropy we also find a weak dichotomy: there are transformations with large time scales that have low central cooling times and, vise versa, there are fast transformations with large central cooling times.  

In summary, according to TNG-Cluster, core transformations exhibit a broad range of timescales, which are comparable to dynamical timescales, with faster transformations ($<2$\,Gyr) at higher redshift and slower ones ($>2$\,Gyr) only at $z<1$. Transformations to higher core entropy are mostly fast (1-2 Gyr) whereas those to lower core entropy seem to be either fast or slow: this may suggest that there are different processes responsible for those transformations.

\begin{figure}
    \centering
    \includegraphics[width=0.46\textwidth]{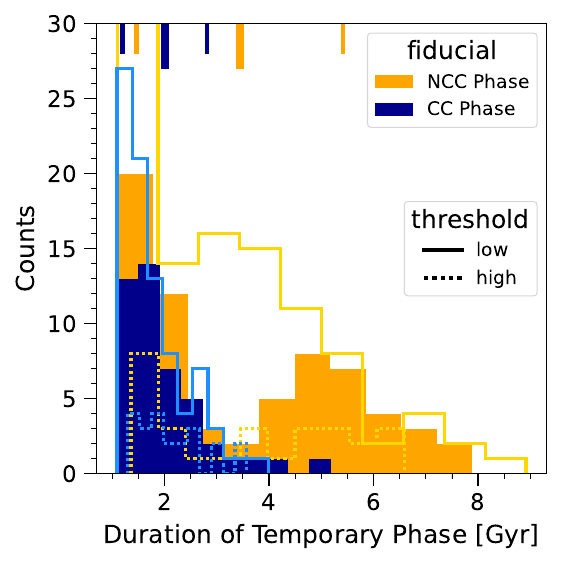}
     \caption{The duration of temporary transformation phases, defined as a period of time between two transformations with opposite derivative signs: namely, one transformation to lower $K_0$ followed by a transformation to higher $K_0$, or vice versa. We label cool-core phases (CCPs) as those whose temporary core entropy value is lower, and non-cool-core phases (NCCPs) as those with a higher temporary entropy. CCPs are shown in blue, and NCCPs in orange. The short vertical lines at the top of the histogram mark the mean and standard deviation of the distributions. For comparison, the solid (dotted) line histograms show a different and more inclusive (restrictive) selection criteria for transformations. Overall, CCPs are typically shorter than NCCPs, and the distribution of durations for the NCCPs is suggestively bimodal.}
    \label{fig:HistDuration}
\end{figure}

\subsection{The case of temporary phases}
After a core transformation, clusters can occasionally revert and return to their earlier CC or NCC state. In TNG-Cluster, 22\% of the halos have core transformations in both directions, i.e. their transformation phases are temporary. An analysis of these temporary transformations phases, i.e. of pairs of transformations with opposite signs of their derivatives (Fig.~\ref{fig:HistDuration}) reveals, there are 111 of such phases across the TNG-Cluster sample, where 45 are phases that start with a transformation to lower core entropy and end with a transformation to higher core entropy (i.e. a cool-core phase, CCP) and 66 are phases with the opposite order (i.e. a non-cool-core phase, NCCP). 

We find that the distribution of CCPs is single peaked with a mean of 2.0 Gyr, while the distribution of NCCPs is doubled-peaked with mean of 3.4 Gyr. On average, CCPs are shorter than NCCPs i.e. clusters remain temporarily in low core entropy states for shorter times than at high core entropy. 
Furthermore, we find that intermediate duration NCCPs are weaker than shorter and longer ones, and that the distribution of temporary phase durations depends quantitatively on the adopted threshold.
Shorter temporary phases tend to occur slightly closer to $z=0$ than longer ones (not shown), with average starting times of $4.5$\,Gyr (CCP) and $6.5,$\,Gyr (NCCP) ago. Short phases are found across all halo masses, while long duration NCCPs occur only in the most massive halos, that live long enough to experience a long NCC phase, i.e. have sufficient evolutionary time with an appropriate merger/feedback history for their NCC to eventually cool and return to a CC state.

The results of Fig.~\ref{fig:HistDuration} may suggest that CCs are more easily disrupted, while NCCs either require more time to transition back to a CC state or that the process of returning to a CC can be interrupted. 

{\renewcommand{\arraystretch}{1.3}
\begin{table*}
    \centering
    \caption{Summary of key characteristics of the six archetypal cluster histories across the TNG-Cluster population. For each category we list: the number of halos of that class and then the averages across the clusters in each category of: number of transformations experienced by each cluster, formation redshift, halo mass at $z=0$, central entropy at $z=0$, and number of all mergers experienced by each cluster. We give the mean, and in parentheses the median, of each category. In descending order, the six archetypes are clusters with (i) no transformations, (ii) only transformations to higher core entropy, (iii) only transformations to lower core entropy, (iv) transformations to lower, then to higher core entropy (v) transformations to higher, then to lower core entropy, and (vi) clusters with three or more transformations that do not fit into the other classes.}
    \begin{tabular}{lllllll}
        \hline\hline
        Type of history & \# halos & \# transformations & $z_{\rm{form}}$ & $\log(\mvir/10^{14}\msun)$ &  $K_0$ [keV cm$^2$]& \# mergers\\
        \hline
        No changes                & 35 (10$\%$)     & 0       & 0.90 (0.85) & 3.95 (3.35) & 56 (47)   & 8.2 (6) \\
        only $K_0$$\nearrow$      & 226 (64$\%$)    & 1.6 (1) & 0.94 (0.89) & 5.28 (4.35) & 159 (129) & 11.9 (11) \\
        only $K_0$$\searrow$      & 13 (4$\%$)      & 1.2 (1) & 1.03 (1.04) & 3.29 (2.71) & 20 (15)   & 6.4 (6) \\
        $K_0$$\searrow$$\nearrow$ & 14 (4$\%$)      & 2       & 1.43 (1.36) & 6.49 (5.89) & 87 (31)   & 12.2 (12) \\
        $K_0$$\nearrow$$\searrow$ & 28 (8$\%$)      & 2       & 1.12 (1.41) & 4.82 (3.98) & 44 (26)   & 11.4 (11) \\
        Irregular                 & 36 (10$\%$)     & 3.4 (3) & 1.26 (1.18) & 6.31 (5.93) & 123 (81)  & 13.8 (13) \\
        \hline\hline
    \end{tabular}
  \label{tab:archetypes}
\end{table*}}

\begin{figure*}
    \centering
	\includegraphics[width=0.48\textwidth]{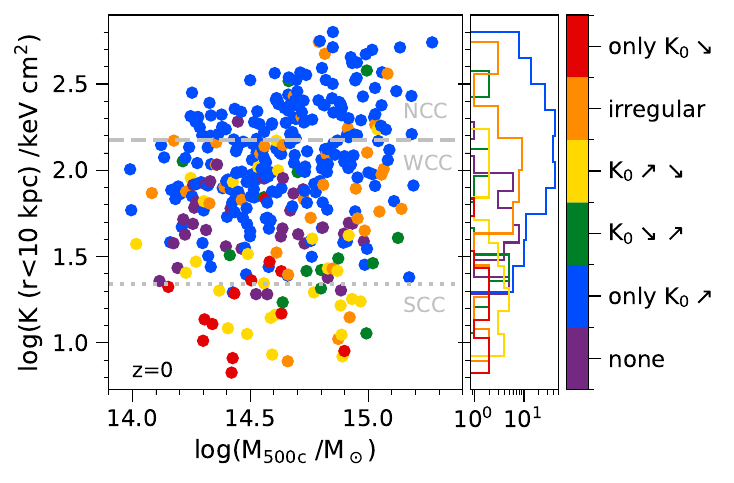}
    \includegraphics[width=0.48\textwidth]{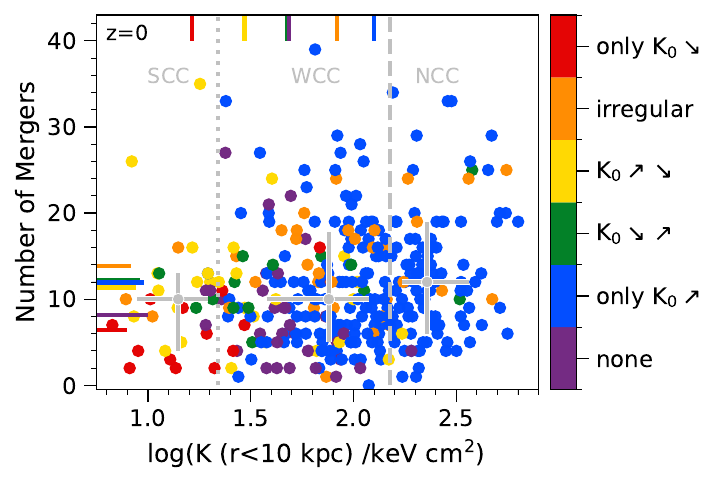}
     \caption{Correlation of central entropy at $z=0$ with the cluster mass at $z=0$ (left panel) and numbers of mergers (right panel), for all halos in TNG-Cluster. Each cluster is colored by its archetype (see text and Table~\ref{tab:archetypes}). The gray horizontal or vertical lines indicate the central entropy values used to classify (N)CCs and the gray markers in the right plot indicate the mean number of mergers and the 1$\sigma$ range for SCCs, WCCs and NCCs. The subplot in the left plot shows the marginalized histograms of central entropy for each of the six categories. We find that clusters experiencing only changes to higher core entropy (blue) end up with higher core entropy at $z=0$ and are found across the whole mass range. In contrast, clusters that only change toward lower core entropy (red) are preferably found among lower mass clusters. In addition, the average number of mergers for the latter case is smaller than for clusters with changes to high core entropy only, suggesting that mergers likely play a role in driving transformations.}
    \label{fig:K0vsM500vsTrafoClasses}
\end{figure*}

\subsection{A diversity of past histories for the cores of clusters} 

As the figures and results presented above demonstrate, according to TNG-Cluster the cores of clusters may undergo a large diversity of thermodynamical evolutions. 
We propose a qualitative classification of TNG-Cluster halos into six distinct archetypes based on their evolutionary paths, and examine the characteristics of these six categories. We show the evolution of central entropy for exemplary halos of each category in Fig.~\ref{fig:scenarios}.

The average properties of these archetypes are given in Table~\ref{tab:archetypes}. In our sample 35 (10$\%$) clusters have no transformations. As the category with the most members, we find 226 (64$\%$) halos with changes to higher core entropy only ($K_0\, \nearrow$), i.e. in the qualitative direction from CCs to NCCs. On average each cluster in this category experiences 1.6 transformations (values range from 1 to 5). In contrast, we identify only 13 (4$\%$) clusters with changes to lower core entropy only ($K_0\, \searrow$), i.e in the qualitative direction from NCCs to CCs. Each has 1.2 transformations on average (i.e. between 1 to 2 per halo).

We also identify two categories of cluster histories, totaling to 12\% among all, that contain two distinct transformations: one where the core entropy increases before decreasing (28, $8\%$; $K_0\,  \nearrow \, \searrow$), and vice versa (14, $4\%$; $K_0\,  \searrow \, \nearrow$). For 20$\%$ of these clusters, the central entropy before and after the two transformations differs only by 10$\%$. Finally, the last category comprises all clusters with three or more transformations; we call these cases `irregular clusters'. There are 36 ($10\%$) halos in this group, and they each undergo on average 3.4 changes since they reached a mass of $10^{14}~\msun$.

Clusters without transformations, and with changes to higher core entropy only, form latest, while irregular clusters and $K_0\,  \searrow \, \nearrow$-clusters form the earliest. Those two types of cluster archetypes are also the most massive halos by $z=0$. As expected, clusters with transformations to higher core entropy only end up with the highest central entropy today, while halos that experience only changes to lower core entropy have today the lowest entropy.

To explore these evolutionary histories as a function of cluster properties, Fig.~\ref{fig:K0vsM500vsTrafoClasses} shows the dependence of mass and central entropy on the cluster archetype at $z=0$ (left panel). Halos that experience changes to higher core entropy only (blue) are found across all masses, but have large $z=0$ central entropy: at $z=0$ all of the clusters of this category are classified as WCCs or NCCs. In comparison, clusters with transformations in the opposite direction only (red) have preferably lower masses and are mostly classified as SCCs at $z=0$.

Clusters with no transformations (purple) are mainly classified as WCCs and have a large scatter in halo mass. Of the 35 cluster in that class 3 are born as SCCs (with values very close to the WCCs threshold) and the rest is born as WCCs. $K_0\,  \nearrow \, \searrow$ ($K_0\,  \searrow \, \nearrow$) halos are shown in yellow (green). The $K_0\,  \nearrow \, \searrow$ clusters are preferably found in the WCC and SCC regime, but are scattered across all masses. $K_0\,  \searrow \, \nearrow$ clusters are scattered across all entropy values, but are more often found at higher masses. Finally, irregular clusters (orange) exhibit a larger scatter in mass and central entropy and are found across the parameter space. 

Importantly, clusters that are SCC today are mainly those systems that either (i) experience changes to lower core entropy only, or (ii) are among $K_0\,  \nearrow \, \searrow$ halos. In contrast, NCC clusters are predominantly those that experience changes to higher core entropy.

Our manual classification of the evolutionary histories of cluster core entropy clearly reflects trends and sub-populations with distinct physical properties. However, we emphasize that while these classes capture overarching trends, the vast majority of clusters experience only transformations to higher core entropy.


\section{Mergers and cluster core transformations} \label{sec_mergers}

We begin to study the underlying causes of core transformations by evaluating the role of mergers. In particular, we start by exploring the merger histories of our six archetypal classes.

Given our merger definitions (Sec.~\ref{sec_defMerger}), we identify a total of 4054 mergers (secondaries when crossing $0.5\rvir$ of the primary halo) for the 352 TNG-Cluster systems since their formation redshift, i.e. since they exceeded the $10^{14}~\msun$ mass threshold. On average each cluster has 11.5 such mergers. These numbers apply to the entire TNG-Cluster population considering the lifetime of the clusters after their formation. 

Fig.~\ref{fig:K0vsM500vsTrafoClasses}, right panel, shows the numbers of mergers experienced by each cluster, as a function of their central entropy at $z=0$. Correspondingly, the last column of Table~\ref{tab:archetypes} gives the average number of mergers for each cluster for the six classes. In both cases, we include all mergers with mass ratios above 0.01, but the same trends also hold for major mergers alone. 
Overall we can say that the clusters that underwent more mergers eventually have the highest core entropy at $z=0$, even though the bulk of the TNG-Cluster population exhibits similar number of past mergers across a wide range of $z=0$ core entropy values. In other words, the number of mergers is not a predictor of $z=0$ cool-coreness (large gray crosses of Fig.~\ref{fig:K0vsM500vsTrafoClasses}, right panel), or rather it is only in the case of clusters that undergo a significantly larger number of mergers than the average. 

Importantly, however, there are differences in merger histories across the identified archetypal classes. Clusters with no transformations experience only 8.2. mergers on average. The archetype $K_0\, \searrow$, i.e. those that move towards low entropy only, has a similarly low number of mergers, i.e. half of the average (6.4), while the archetypes $K_0\, \nearrow$, $K_0\,  \nearrow \, \searrow$, and $K_0\,  \searrow \, \nearrow$ (which dominate the sample) have on average 11.9, 11.4, and 12.2 mergers, as per the global average. Irregular clusters experience the largest number of mergers (13.8). Together, these findings suggest a significant connection between merging events and thermodynamical transformations of cluster cores. That is, mergers may drive some or even the majority of transformations, as we further quantify below. 

\subsection{Correlation between mergers and core transformations}

\begin{figure}
    \centering
    \includegraphics[width=0.46\textwidth]{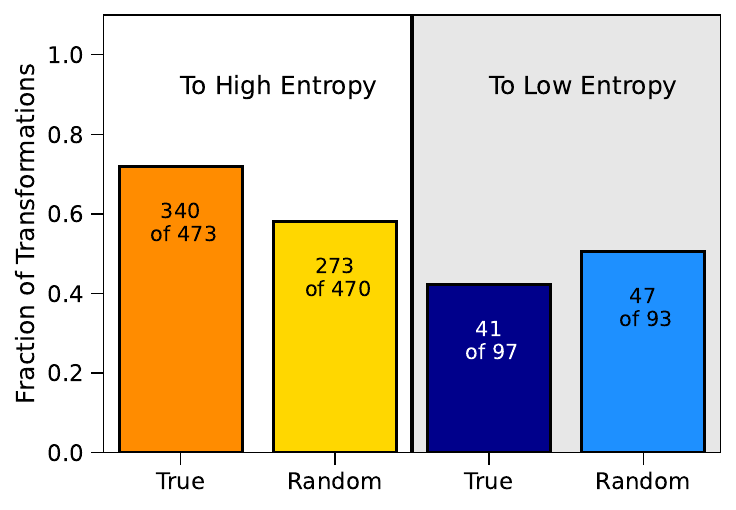}
     \caption{Fraction of transformations with at least one merger occurring within 1 Gyr prior to the transformation according to TNG-Cluster. The left half of the figure shows the respective numbers for transformations to higher core entropy, while the right half shows the fractions for transformations to lower core entropy. To put these fractions into context, we present for each type of transformation the fractions of a random sample. (See text for details on random sample generation). We find that transformations to higher core entropy are more likely preceded by a merger.}
    \label{fig:fracTwwoM}
\end{figure}

Are mergers and core transformations correlated in time? Fig.~\ref{fig:fracTwwoM} shows the fraction of transformations that are preceded by a merger within a time window of 1 Gyr. We find that 72$\%$ of all transformations to higher core entropy are associated with a merger. That this fraction is not 100\% suggests that either (i) there are additional processes that causes transformations, which are responsible for the other $\sim 1/3$, and/or that mergers may induce transformations on timescales longer than a Gyr. On the other hand, only 42$\%$ of transformations to lower core entropy have a temporally-close merger. If anything, mergers seem to suppress the transition of clusters from more NCC to more SCC states.

To provide context, we compare both fractions to random samples. To do so we shuffle the merger histories of halos: for each cluster, we consider its transformations, but take the mergers from a different randomly selected halo. This process is repeated 100 times. For transformations to lower core entropy, it is similarly -- even somewhat less likely -- to find a recent merger, compared to the random sample. In contrast, for transformations to higher core entropy, it is more likely to have a recent merger than in the random sample.

\begin{figure}
    \centering
    \includegraphics[width=0.48\textwidth]{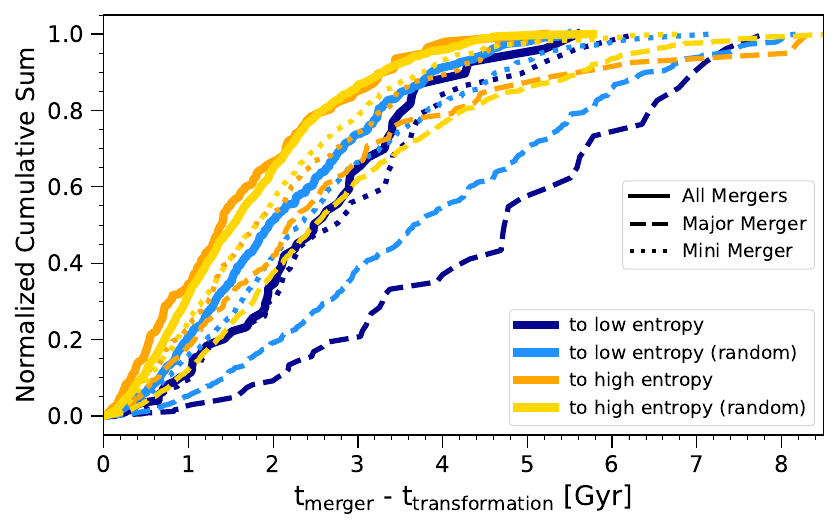}
	\includegraphics[width=0.48\textwidth]{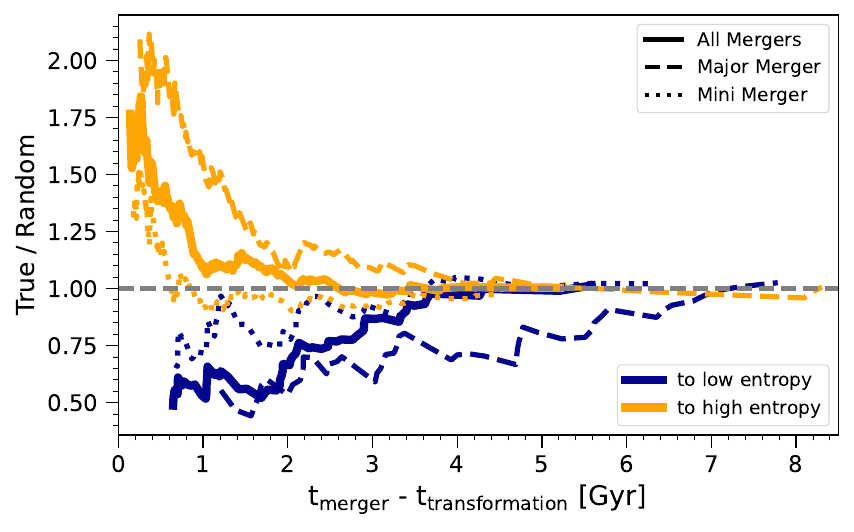}
    \caption{Top panel: cumulative sum of the time difference between each transformation and its closest merger for all TNG-Cluster halos. We separate transformations to higher (orange) and lower (dark blue) core entropy, along with the corresponding random samples (gold and light blue). Bottom panel: ratio of the true and random samples for each type of transformation from above. Overall, mergers are significantly more likely found within 1\,Gyr before a transformation to higher core entropy than in the random sample. In contrast, it is less likely to find a transformation to lower core entropy up to 3.5\,Gyr after a merger. This indicates that changes to higher core entropy are strongly correlated with mergers, while changes to lower core entropy may even be anti-correlated. This connection is strongest for the smallest time differences, and much stronger for major mergers (dashed lines) than mini mergers (dotted lines).}
    \label{fig:cumHistdtMerger}
\end{figure}

To measure the time scale of possible association between mergers and transformations, Fig.~\ref{fig:cumHistdtMerger} shows the cumulative sum of the time differences between the closest pairs of transformations and mergers. We assign such pairs by looping though all transformations of a halo (from large to small lookback times) and finding the closest merger for each.\footnote{We assign mergers only once, implying that some transformations cannot be associated with a merger. For transformations to lower core entropy this never happens. Vice versa, only 6 transformations are not assigned a merger.} Transformations to higher core entropy (orange) exhibit a steep increase at small time differences, and flatten out at large time differences. In contrast, transformations to lower core entropy (dark blue) mildly increase at first, and only exhibit a steeper slope for larger time differences. This indicates that the median time difference between mergers and transformations to higher core entropy (0.3\,Gyr) is smaller than for transformations in the opposite direction (1.1\,Gyr) and that there are more pairs with small time differences for changes to higher core entropy than vise versa. For transformations to higher (lower) core entropy the line of the true sample is above (below) the shuffled, randomized sample, indicating significance above random.

To quantify this significance, we divide the true distributions by the random distributions (Fig.~\ref{fig:cumHistdtMerger} lower panel). We show lines only for $N \geq 7$ in the cumulative sums, to avoid noise at small times. A clear excess is present for transformations to higher core entropy (orange lines) within 1\,Gyr. This is indicated by a ratio above unity. The signal is stronger, and persists to longer timescales, for major mergers (dashed), while it is present only at the smallest time separations for mini mergers (dotted). In contrast, transformations to lower core entropy are always less likely to occur than random, with ratios below unity. This signal persists to 3.5\,Gyr following a merger. Collectively, this suggests a strong positive correlation between mergers and cluster core transformations towards higher entropy. Simultaneously, it hints towards an anti-correlation for transformations towards lower entropy, i.e. mergers inhibit the cooling and settling of cluster cores. The strength of both trends are highest for the shortest time differences between associated mergers and transformations, as would be expected.

\begin{figure}
    \centering
    \includegraphics[width=0.48\textwidth]{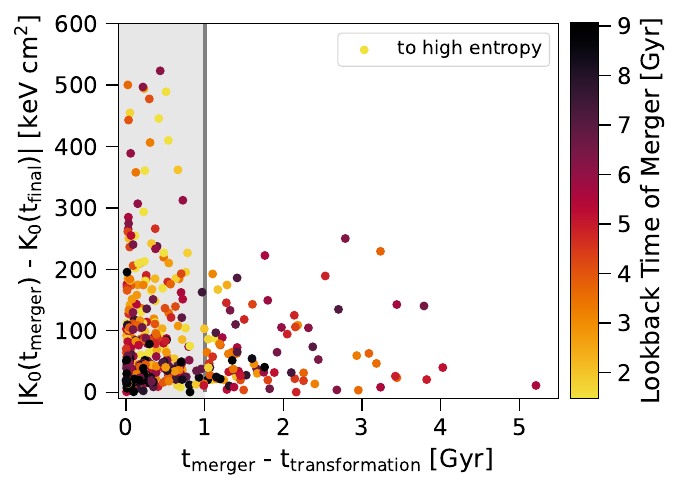}
    \caption{Strength of change of core entropy (y-axis) as a function of the time difference between the closest merger and the transformation. The change in entropy $\rm{d}K_0$ is measured between the time of the merger and end of the transformation for all halos in TNG-Cluster. Color indicates the lookback time of the merging event. We find that large entropy changes occur only for small time differences, suggesting that mergers identified within $\sim 1$\,Gyr prior to a transformation are physically associated and/or drive strong cluster core evolution, while mergers that occur earlier than this time window are unassociated and/or cause weaker changes in core entropy.}
    \label{fig:dtVsDK0VsLbtMerger}
\end{figure}

Finally, we quantify the possibly causal impact of mergers on core transformations to higher core entropy by measuring the change in cluster core entropy in a core transformation as a function of time since the preceding merger. Fig.~\ref{fig:dtVsDK0VsLbtMerger} shows $\rm{d}K_0 \equiv K_0(t_{\rm final}) - K_0(t_{\rm merger})$ where $t_{\rm final}$ is the time marking the end of the transformation, and $t_{\rm merger}$ is our same definition of merger time. Larger values of $\rm{d}K_0$ indicate stronger transformations. We find that the largest $\rm{d}K_0$ values are found only for merger-transformation pairs with small time differences. In particular, the rapid drop of large entropy jumps for time-scales greater than $\sim$\,1\,Gyr motivates our earlier use of this threshold. In addition, these merger events preferentially occur at small lookback times i.e. near $z=0$, as indicated by the colorbar. High-redshift mergers produce smaller core entropy increases.

A few additional details and caveats should be kept in mind. First, these quantitative results depend on the adopted merger definitions, for the mass ratio as well as the merger time. When we instead define a merger mass ratio at the time of the crossing of $0.5 r_{\rm 500c}$, we find a stronger correlation with changes to higher core entropy persisting to similar timescales (lower panel in Fig.~\ref{fig:ap_cumhist_variedMRrCross}). The anti-correlation of mergers and changes to lower core entropy is of similar strength but persists to larger timescales (up to 6\,Gyr). At face value, the correlation of the entire population with the mass ratio defined at the crossing point closely resembles the trends observed for major mergers when the mass ratio is defined at the maximum past mass. This similarity can be attributed to the fact that the mass ratio at maximum mass is generally larger than at the crossing point. Specifically, when applying the same threshold of 0.01 for the secondary mass, fewer mini mergers are included in the sample when the mass ratio is defined at the crossing point. As a result, the sample is more dominated by major mergers from the sample defined by maximum past mass. When we define merger times not as $0.5 r_{\rm 500c}$ crossings, but instead as $r_{\rm 200c}$ crossings, i.e. halo infall times, we find a weaker correlation (anti-correlation) of mergers and changes to higher (lower) core entropy for $r_{\rm 200c}$ crossings (upper panel in Fig.~\ref{fig:ap_cumhist_variedMRrCross}). In addition, when separating mergers into different mass ratio categories, it is important to recall that smaller mergers are much more frequent than major mergers. As a result, it is easy to assign temporally-close mini mergers to transformations, but this does not necessarily reveal any casual relation.

\begin{figure*}
    \centering
    \includegraphics[width=0.44\textwidth]{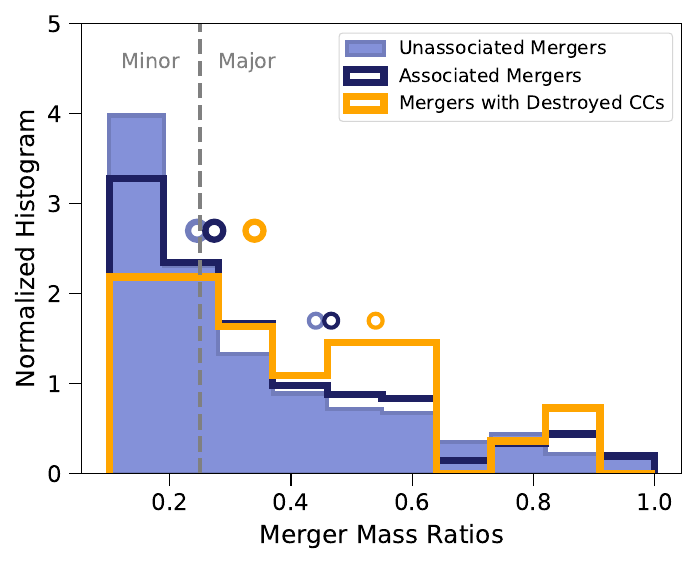} 
    \includegraphics[width=0.44\textwidth]{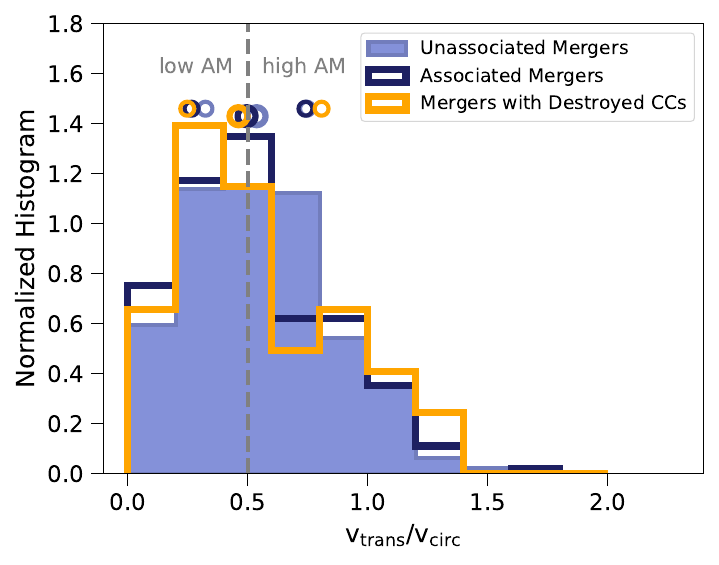}

    \includegraphics[width=0.67\textwidth]{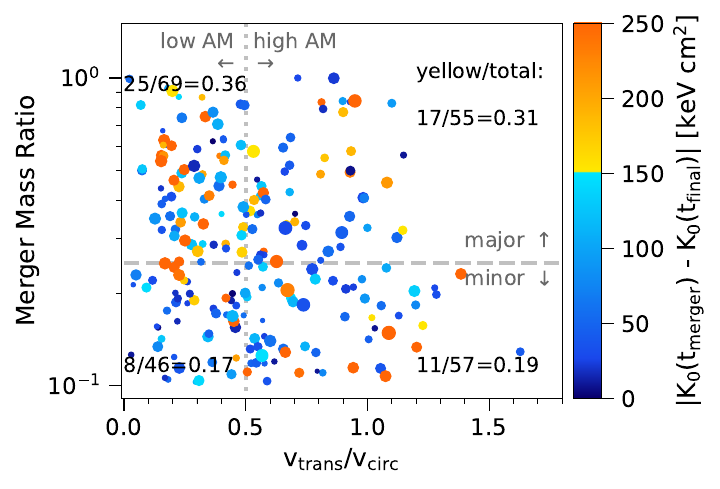}
    \caption{Connection of mergers properties and the destruction of cluster cores in TNG-Cluster.
    The upper row shows the distributions of merger mass ratio (left) and tangential velocity (right), for mergers that are not associated with a core transformation (light blue), mergers that are closely paired with a cluster core transformation (dark blue), and mergers that are closely paired with a cluster core transformation that has $K_0(t_{\rm final}) - K_0(t_{\rm merger}) \geq 150 $\,keV\,cm$^2$ (which we dub as cases where the CC is destroyed, orange). The distributions are somewhat skewed towards higher mass ratios and lower angular momentum for associated cases, suggesting that these scenarios preferentially lead to cluster core transformations. The shift is slightly stronger for transformation merger pairs with large $\rm dK_0$. The lower plot shows the correlation of merger mass ratio and tangential velocity of all closely associated major and minor mergers $(\rm{dt} < 1\,\rm{Gyr})$ and the possibility that a CC may be disrupted. The color indicates the change in entropy, where orange shows large changes in entropy $\geq$\,150\,keV whereas small changes $<$\,150\,keV are in blue. We have chosen a threshold of 150\,keV to be consistent with the fact that NCCs are defined by $K_0 > $\,150\,keV. The size of the dots gives the main halo mass at the time of the merger. The text label in each quadrant gives the number of large changes, divided by the total number, in that quadrant, as well as the resulting fractions. We find that large changes i.e. destroyed cool cores are most abundant among the major mergers with low angular momentum, but most importantly cool cores may be destroyed also otherwise. }
    \label{fig:HistMR}
\end{figure*}

Amid unavoidable complexity and details, we can conclude that the tests and figures above demonstrate that, according to TNG-Cluster, there are strong indications of time correlations between merger and core transformation events: this is the case exclusively for transformations towards higher core entropy, whereas transformations towards lower core entropy are not statistically correlated with preceding merger events. 
In fact, we have specifically identified mergers that precede core transformations, given that they could only influence cluster thermodynamics in the future. However, it is possible that, although time correlated, mergers are not the dominant cause of core transformations, or not the direct cause for transformation. For example, they could trigger an intermediate physical process, such as strong AGN feedback, which could then directly drive transformation. We consider this possibility in more detail below, after a few more inputs on the types of mergers.

\subsection{Properties of mergers associated to core transformations}

Are a subset of mergers responsible for cluster transformations towards high entropy? If so, what properties differentiate these merger events from the general population? We focus on the merger mass ratio and angular momentum (as described in Sec.~\ref{sec_defMerger}), motivated by previous findings \citep{poole2008, hahn2017, valdarnini2021, chen2024}.

Fig.~\ref{fig:HistMR}, top left, shows the distribution of merger mass ratio separated by merger events that are (closely) associated to a transformation event (dark blue), versus those that are not associated with a transformation (light blue). This pairing is based on our fiducial time window threshold of $t_{\rm merger}-t_{\rm transformation} < 1\,$Gyr. We see that both histograms are dominated by mergers with the smallest mass ratios. The dark blue distribution, i.e. the one mergers that can be associated to core transformation, is skewed towards somewhat larger values, with larger median and 75th percentile of the distributions: 0.047 and 0.15 for unassociated mergers, versus 0.057 and 0.24 for associated mergers. These findings suggest that major mergers may be more tightly correlated with transformations to higher core entropy than mini or minor mergers. However, the magnitude of this effect is really small.

We also evaluate differences in the angular momentum (AM) of mergers. Fig.~\ref{fig:HistMR}, top right panel, shows the ratio of tangential to circular velocity for mergers associated with transformations (dark blue), and those that are not (light blue). As above, the differences between these two distributions are subtle. The median is 0.60 (light blue) vs 0.58 (dark blue), while the 25th and 75th percentiles are shifted from 0.38 and 0.83 (light blue) to 0.37 and 0.80 (dark blue). We also note that when we repeat the analysis of Fig.~\ref{fig:cumHistdtMerger}, evaluating the time difference between mergers and transformations in comparison to a control sample, mergers with low AM show a stronger excess than those with high AM (see Fig.~\ref{fig:ap_cumhist_3vt}). Taken together, these findings suggest that low AM encounters may be more effective in changing the core entropy, however the connection is rather weak. This could arise if the mergers that transform cluster cores are not special with respect to either mass ratio or AM, or that the dependencies therein are more complex.

Finally, we consider whether the strength of the entropy change $\rm{d}K_0$ depends on merger mass ratio or angular momentum. We find, although do not show, that mergers with the largest $\rm{d}K_0$ do correspond to low angular momentum and/or high mass ratio encounters.  We go even further by combining the insights from Fig.~\ref{fig:dtVsDK0VsLbtMerger} and the upper panels in Fig.~\ref{fig:HistMR} to investigate which types of mergers may disrupt a cool core and whether this process depends on mass ratio and/or angular momentum. We assume that a cool core is destroyed if the change in entropy is larger than 150\,keV\footnote{We choose this threshold because NCCs are defined by $K_0 > $ 150\,keV, noting that this is a rather conservative threshold that excludes some fairly strong transformations.}. 

In the upper panels of Fig.~\ref{fig:HistMR} orange histograms show the distributions of mass ratios (left) and AM (right) for the mergers associated with the destruction of a CCs, i.e. with the large $\rm{d}K_0$, to be compared to the other dark and light blue histograms. Mergers associated to destroyed CCs have typically larger mass ratios and, to a lesser degree, more radial (i.e. lower AM) encounters.

The lower plot in Fig.~\ref{fig:HistMR} therefore shows the correlation of angular momentum and mass ratio of closely correlated mergers and transformations ($\rm{dt}<1$\,Gyr). We include only minor and major mergers, excluding mini mergers with mass ratios $< 1/10$. The color of the dots indicates the change in entropy $\rm{d}K_0$: strong transformations that could destroy a cool core are indicated in orange, while weak transformations are shown in blue. Marker size reflects halo mass at merger time, and we see that more massive clusters have generally stronger increases in central entropy. We also examine the trend with lookback time, and do not find any correlation across this plane, indicating that the lookback time of the merger is not directly relevant (not shown).

We divide the plane into four quadrants that trace high and low mass ratio and angular momentum mergers, respectively. The text labels provide the fraction of strong to total transformations in each quadrant. We find that the most occupied quadrant is the one with low angular momentum and high mass ratios (upper left). Conversely, the least occupied quadrant has low angular momentum and low mass ratios (lower left). Importantly for the discussion at hand, the former also has the highest fraction of strong transformations. This supports our and others previous findings that major merger with low angular momenta are most capable of destroying a cool core. We also see that major merger are more likely to destroy a core than minor mergers, whereas the trend with angular momentum is less clear. According to TNG-Cluster and our analysis, the mass ratio seems to be a stronger predictor for the disruption of a cool core than angular momentum.

In fact, Fig.~\ref{fig:HistMR} shows that, according to TNG-Cluster and across the sample, it is possible to identify mergers that destroy cool cores with \textit{all possible} angular momenta and mass ratios. The idea that only major mergers, or only mergers with strongly radial orbits, can disrupt CC clusters is incorrect. The connection between mergers and core transformations is more complex, potentially due to dependencies on additional factors beyond those we have explored. Mergers are also frequent and relentless and, even when a single event does not disrupt the cluster core, a repeated series of (possibly smaller) mergers and their cumulative injection of energy into the core could potentially do so.

\begin{figure*}
    \centering
    \hspace*{-1.8cm}\includegraphics[width=0.9\textwidth]{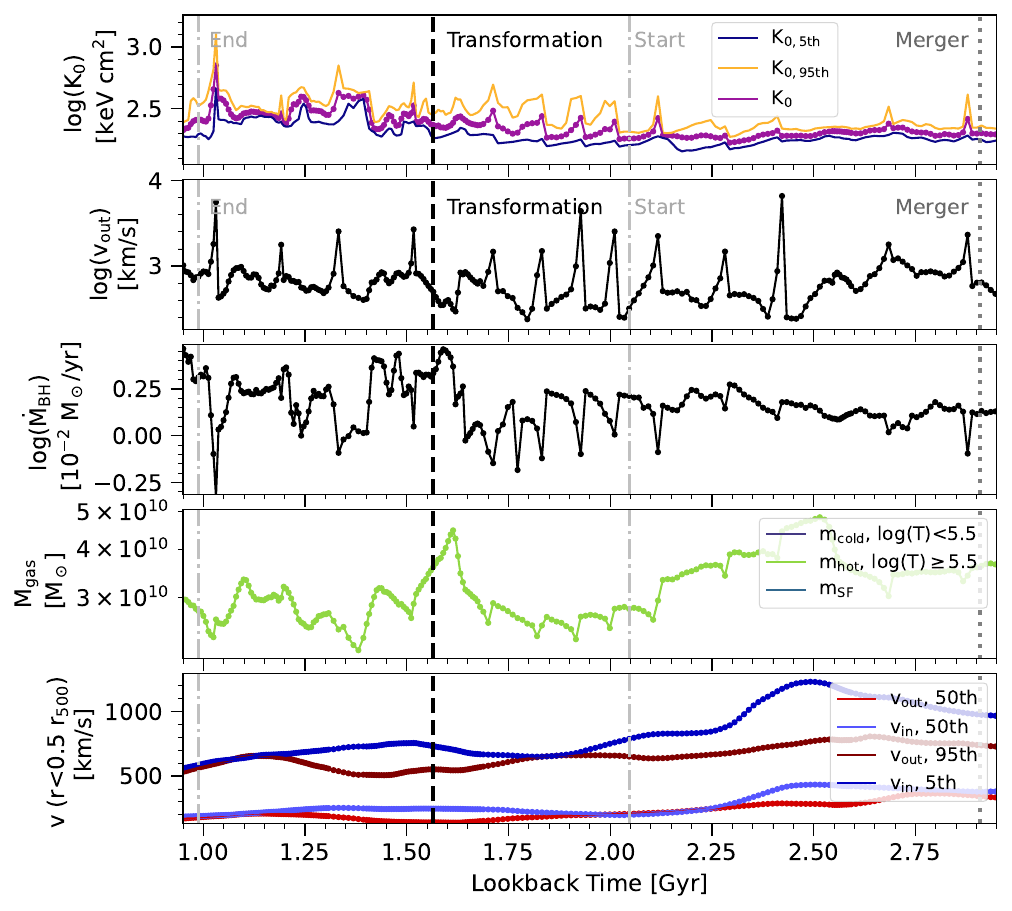}
    \caption{Evolution of the thermodynamical cluster core properties, as well as AGN feedback activity, as a function of time of an example cluster, the most massive cluster of TNG300 for which we have high temporal cadence output. We show a single halo, followed using the high time-resolution subbox (see text). We focus on a $\sim 2$\,Gyr period of time centered on $z \simeq 0.1$ and the last transformation, and last merger, of this cluster. From top to bottom the panels show: the mean central entropy measured within 10 kpc (purple) and the 5th (blue) and 95th (orange) percentiles; the 95th percentile of the outflow velocity within 10 kpc around the central black hole (black); the instantaneous accretion rate of the central black hole (black); the gas mass within 30 kpc split into hot ($T\geq 10^{5.5}$ K, green), cold ($T< 10^{5.5}$ K, blue), and star forming phases (teal); the bulk outflow and inflow velocities within $0.5\rvir$ (red and blue, respectively) and the 5th percentile of the inflow velocity (dark blue) and the 95th percentile of the outflow velocity (dark red) also measured within $0.5\rvir$. The vertical black dashed lines indicate the transformations, the light gray dash-dotted lines indicate the start and end of a transformation, and the dark gray dotted lines show the time of the closest associated merger (all lines are labeled in the first two rows). Repeated AGN feedback episodes are visible as short time-scale peaks in outflow velocity and central entropy that collectively help to transform this cluster into a NCC state.}
    \label{fig:timeseries}
\end{figure*}

To summarize, our analysis reveals several correlations between mergers and cluster core transformation events. Mergers often take place just prior to transformations to higher core entropy, particularly within $\sim$\,1\,Gyr. The occurrence rate is more than expected from random association, and is more prominent for major as opposed to minor mergers. Mergers associated with core entropy transformations have larger mass ratios and, to a lesser degree, lower angular momentum i.e. more radial orbits, although these trends are weak. Transformations that occur within $\sim$\,1\,Gyr of a merger result in substantially larger increases in central entropy $K_0$ than otherwise, and major mergers are more likely to destroy CCs than minor mergers. These correlations suggest, although do not prove, a physical and thus causal link. Namely, that (at least some fraction of) mergers appear to disrupt cluster cool cores (to some degree), increasing their central entropy, and leading to final states that are more NCC in their physical properties.


\section{AGN feedback-driven cluster core transformations} \label{sec_agn}

\begin{figure*}
    \centering
    \hspace*{-1.8cm}\includegraphics[width=0.9\textwidth]{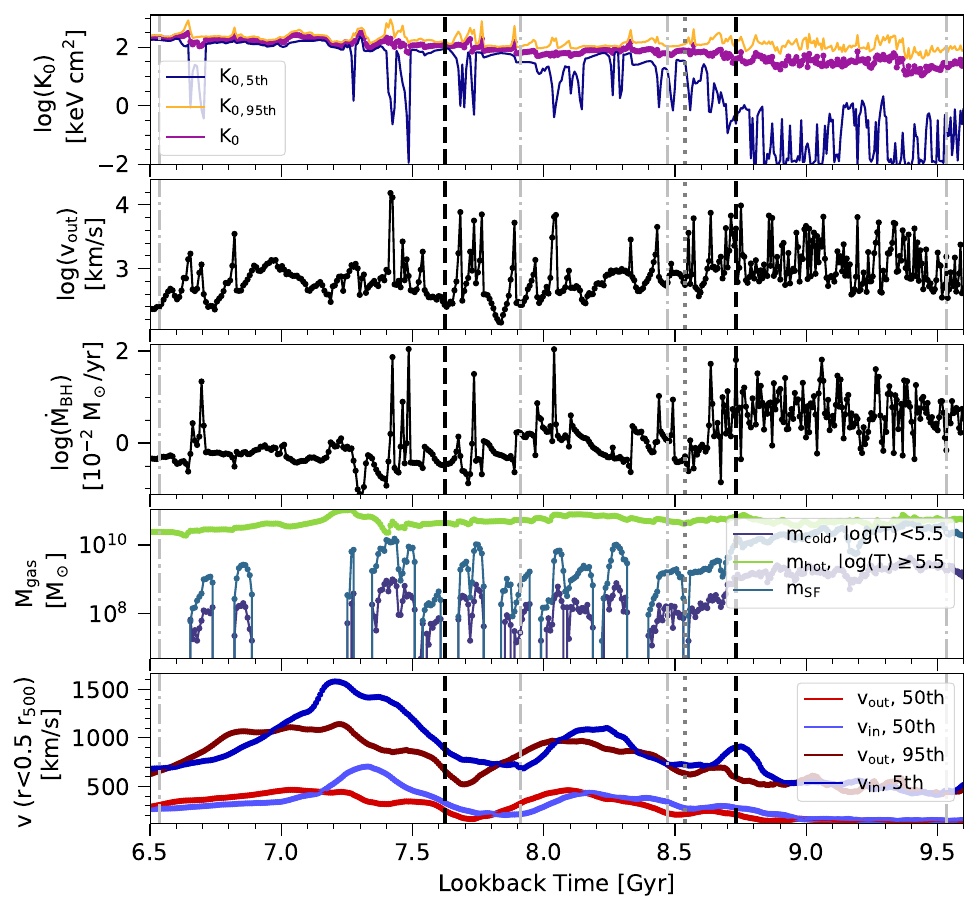}
    \caption{As in Fig.~\ref{fig:timeseries}, showing the same five quantities and tracking the same cluster from TNG300 through time but here focusing on high redshift, i.e. showing details of the first two transformations to higher core entropy. These occur at $z \simeq 1.2$ and $z \simeq 0.95$, and are marked with the vertical black dotted lines. The merger at $t_{\rm lb} \simeq 8.5$\,Gyr causes strong bulk motions in the inner ICM, enhancing the episodic AGN accretion and feedback, that collectively transform and remove cool gas from the cluster core.}
    \label{fig:timeseries2}
\end{figure*}

What impact does the AGN have on cluster cores, and can AGN feedback disrupt cool-cores? We now proceed to connect energy injection by the central AGN to the previously identified core transformations. We do so by neglecting, at first, the presence of mergers, whereas we discuss the possible interplay between mergers and AGN feedback in Section~\ref{sec_discussion}.

Given that AGN feedback is a short time-scale, and potentially short duty-cycle phenomenon, we turn to the high time-resolution outputs available in form of simulation subboxes. These are available for TNG300- 1 (aka TNG300, with the same physics and resolution of TNG-Cluster) and are spatial cutouts of fixed comoving size within the main simulation box and offer a much higher time resolution ($\sim 5-10$\,Myr) compared to the main box ($\sim 150$\,Myr).

We begin with the first subbox of TNG300-1, which is centered on the most massive halo of that volume, with $M_{\rm halo} \sim 2 \times 10^{15} \msun$ at $z=0$. This is representative of the TNG-Cluster sample, and given that TNG300 and TNG-Cluster have identical physics and resolution, we can treat this dataset as an extension of TNG-Cluster. We study the evolution of this cluster as a characteristic and representative example. We identify transformations and the timescale of the transformations by using the exact same methodology as elsewhere (see Sec.~\ref{sec_findTrafos} and Sec.~\ref{sec_def_timescale}). Mergers are likewise identified with the same definitions as above. 

This halo forms at a lookback time of $10.3$ Gyr, i.e. at $z\sim2$ and undergoes three transformations from lower to higher core entropy, i.e. transformations from CC towards NCC states. These occur at lookback times of $8.7$\,Gyr, $7.6$\,Gyr, and $1.6$\,Gyr (i.e. at $z=1.2,0.9,$ and $0.1$). The halo experiences two major mergers, at lookback times of 2.9\,Gyr and 6.4\,Gyr ($z=0.24$ and $0.7$, respectively), with mass ratios 0.3 and 0.5, and with angular momenta $v_{\rm trans}/v_{\rm circ}$ 1.9 and 0.3, respectively. It also undergoes two minor mergers (at lookback times of 8.5\,Gyr and 9.8\,Gyr, with mass ratios 0.16 and 0.23, and with $v_{\rm trans}/v_{\rm circ}$ 3.8 and 1.0, respectively). Even though this cluster has several mergers preceding core transformations, in this section we focus exclusively on the possible time relationship between SMBH energy injections and core transformations. A discussion about combined effects of mergers and AGN feedback of this prototypical cluster can be found in Sec.~\ref{sec_discAGNMerger}. The halo also undergoes one transformation from higher to lower core entropy shortly after crossing the mass threshold for formation, i.e. $10.1$\,Gyr ago. 

Figs.~\ref{fig:timeseries} and \ref{fig:timeseries2} depict the time evolution of this halo. From top to bottom the panels show: the mean central entropy measured within 10\,kpc (purple) and the 5th (blue) and 95th (orange) percentiles; the 95th percentile of the outflow velocity within 10\,kpc around the central SMBH (black); the instantaneous accretion rate of the central SMBH (black); the gas mass within 30\,kpc split into hot ($T\geq 10^{5.5}$\,K, green), cold ($T< 10^{5.5}$\,K, blue), and star forming phases (teal); the bulk outflow and inflow velocities within $0.5\rvir$ (red and blue, respectively) and the 5th percentile of the inflow velocity (dark blue) and the 95th percentile of the outflow velocity (dark red) also measured within $0.5\rvir$. Vertical lines mark mergers (dark grey) and the moment of transformations (black), together with their start and end times (light grey). 

Fig.~\ref{fig:timeseries} focuses around the time of the last i.e. low redshift transformation. It reveals episodic and short-lived peaks in central entropy, which are also visible in the 95th and 5th percentiles. Further, these peaks in $K_0$ are aligned with strong, high-velocity outflows driven by energy injections of the AGN (second panel). As per the known functioning of the AGN feedback model in TNG-Cluster and TNG300, these are driven by kinetic energy injections from the SMBH with low accretion rates \citep[e.g.][]{nelson2019, pillepich2021, prunier2025}. Outflow velocities reach several thousands of km/s during these episodes. As a result, gas is expelled from the core region, lowering the density in the close vicinity of the SMBH. This decreases the instantaneous accretion rate of the SMBH (third panel) and the gas mass in the core (fourth panel). 

Characteristically, the peaks in central entropy rise steeply in time, but then decline more slowly. Therefore, the central entropy can decrease to its pre-AGN episode value during a subsequent period of low AGN activity, \textit{if} there is sufficient time. Throughout the course of the transformation depicted in Fig.~\ref{fig:timeseries}, however, this is not generally the case. Instead, the high frequency of AGN episodes prevent $K_0$ from returning to its initial value. We interpret this as a signature of AGN feedback contributing to the long-term evolution of core entropy towards higher values, i.e. towards the disruption of the cluster cool-core.

Fig.~\ref{fig:timeseries2} is instead centered around the first two transformations, which occur at higher redshift. As this protocluster is already massive enough to host a high-mass, kinetic mode SMBH, we also see significant episodes of effective AGN feedback activity. These are visible in the correlated peaks of outflow velocity (second row) and SMBH accretion rate (third panel). As in the previous case, during such periods of frequent AGN feedback, the central entropy trends towards higher values with time, being unable to fully return to its pre-AGN burst state.

There is little direct response in the mass dominant hot phase (fourth row). However, this protocluster hosts warm/cool and star-forming gas in its core (blue and purple lines), unlike at later times, where only hot gas is present. At $z=0$, massive TNG-Cluster halos tend to host cool gas predominantly in their outskirts \citep{rohr2024a}, while central cool gas reservoirs are the exception rather than the norm (\textcolor{blue}{Staffehl et al. in prep}). During these two high-redshift transformations, the abundance of cooler gas phases shows strong time modulation that is correlated to AGN feedback episodes. In particular, while there is abundant cool gas before the first transformation i.e. at $t > 9$\,Gyr \citep{rohr2024a}, little cool gas remains within the central 30\,kpc after this transformation occurs. This implies that the first transformation destroyed the cold gas reservoir of the cluster. Subsequently, the quasi-episodic and residual levels of central cool gas correlate with the lower percentile of $K_0$, that reflects larger tails away from the mean. 

The timescales of AGN feedback episodes are set by the available gas supply. In the TNG model, kinetic-mode energy injections are time accumulated, until a sufficient amount of accreted rest-mass energy is available to launch a wind episode. If the local gas density around the SMBH is low, preventing efficient accretion, the time between such AGN feedback episodes will be long. On the other hand, if gas densities are high e.g. due to a recent strong inflow towards the cluster center, the cluster being in a CC state, or because of the compressive effect of a merger, the time between AGN feedback episodes will be much shorter. From Figs.~\ref{fig:timeseries} and \ref{fig:timeseries2} it is clear that this duty-cycle varies significantly as a function of time along the life of any single cluster, and that extended periods of rapid cadence AGN energy injections may contribute to CC disruption.

\begin{figure}
    \centering
    \includegraphics[width=0.99\linewidth]{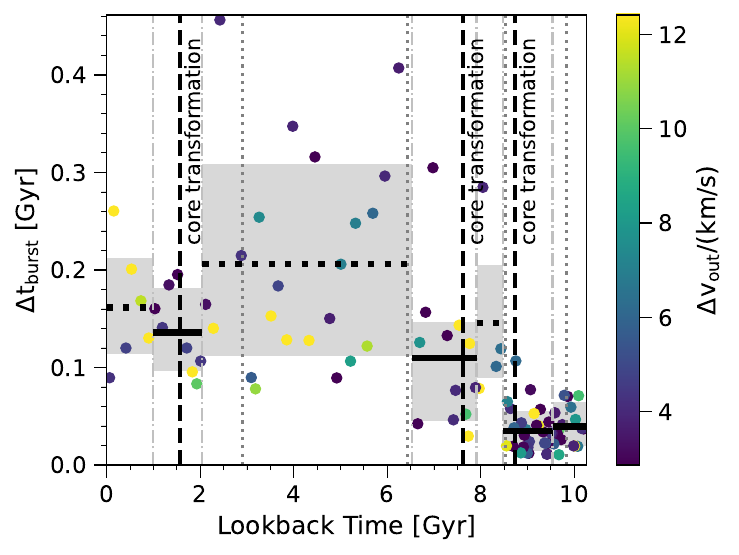}
    \caption{Time between consecutive SMBH feedback episodes, as a function of time and in relation to periods with and without changes of core entropy, for the most massive cluster of TNG300.
    These timings are measured from local maxima i.e. peaks in outflow velocity. Distinct core transformation periods are demarcated with vertical dot-dashed lines, and their times are marked with vertical black dashed lines. Within each core transformation period, the horizontal black lines (gray bands) show the median (one sigma) AGN duty-cycles. Dotted vertical lines indicate mergers (same line styles as Fig.~\ref{fig:timeseries}). Color encodes relative outflow velocity. We see that this halo experiences phases of higher AGN activity (shorter intervals between consecutive energy injections) during all three core transformations, compared to control periods with no core transformation: in the long lull between the early and late time transformation periods ($\sim 6.5$\,Gyr to $\sim 3$\,Gyr) the AGN of this cluster is least active.}
    \label{fig:TimeDiffAGNBursts}
\end{figure}

Fig.~\ref{fig:TimeDiffAGNBursts} quantifies the role of repeated and prolonged SMBH energy injections in destroying CCs by measuring the time difference $\Delta t_{\rm burst}$ between successive peaks in cluster core properties. In particular, we use peaks in the outflow velocity, which are particularly clear and easy to identify and which are a natural proxy for SMBH energy injections. We show the result as a function of lookback time, still focusing on the evolutionary track of the most massive cluster of TNG300. As above, the three core transformation periods are marked by vertical black dotted lines, and their overall duration is bracketed by light gray vertical dash-dotted lines. The merger timings are marked by darker gray vertical lines.

During all three core transformations, $\Delta t_{\rm burst}$ is short, indicating a higher AGN activity during these times. During the first transformation we find $\Delta t_{\rm burst} \sim 30$\,Myr. This value increases to $\Delta t_{\rm burst} \sim 100$\,Myr for the second transformation. Between the second and third transformation the time cadence slows to only $\Delta t_{\rm burst} \sim 200$\,Myr. Finally, during the last transformation the AGN again becomes more active, and this value decreases to $\Delta t_{\rm burst} \sim 130$\,Myr. There is no strong correlation between the strength, that is, the relative height of the peaks, and whether they occur during a core transformation or not (as shown by the color of the markers). This duty-cycle directly correlates with AGN activity: the shorter the time difference, the more energy injected per unit time by the AGN. As a result, AGN feedback has a correspondingly larger ability to influence the thermodynamics of the cluster core during transformation episodes.

Because of the lack of needed output data, we cannot demonstrate the time (and hence plausibly causal) relationship between repeated and prolonged SMBH energy injections and CC destruction for all the halos of TNG-Cluster as for the example cluster of the figures above. However, in Appendix~\ref{sec_tng50}, we give account of the time evolution of the most massive cluster of the TNG50 simulation \citep{nelson2019b, pillepich2019}: all the conclusions above hold also for this cluster, with almost one order of magnitude lower mass than that of TNG300, same galaxy formation physics but at more than one hundred times better mass resolution than in TNG300 and TNG-Cluster.


\begin{figure*}
    \centering
    \hspace*{-0.7cm}\includegraphics[trim={0 0.2cm 0 0},clip,width=1.15\linewidth]{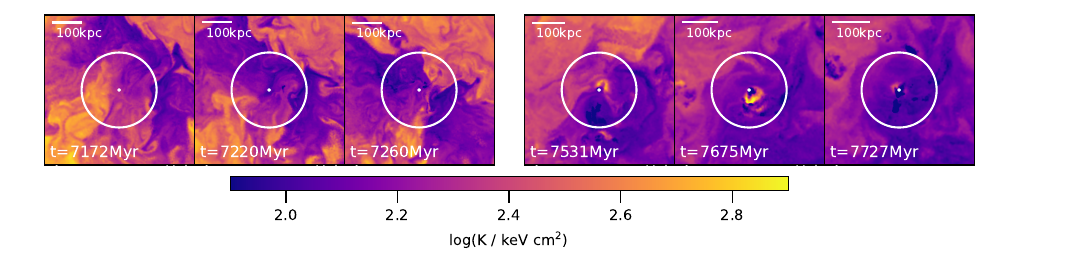}
    \hspace*{-0.7cm}\includegraphics[trim={0 0.2cm 0 0},clip,width=1.15\linewidth]{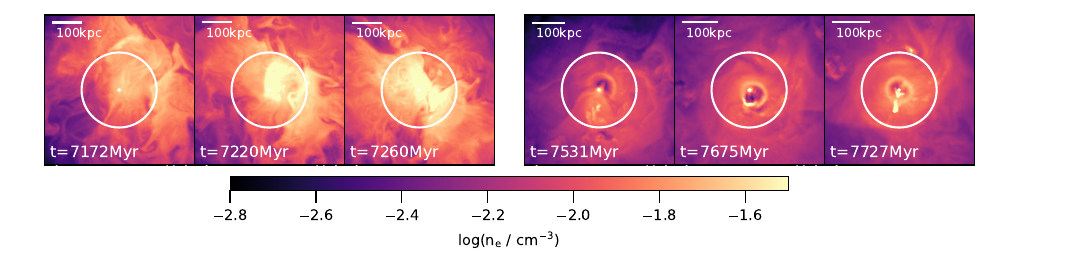}
    \hspace*{-0.7cm}\includegraphics[trim={0 0.2cm 0 0},clip,width=1.15\linewidth]{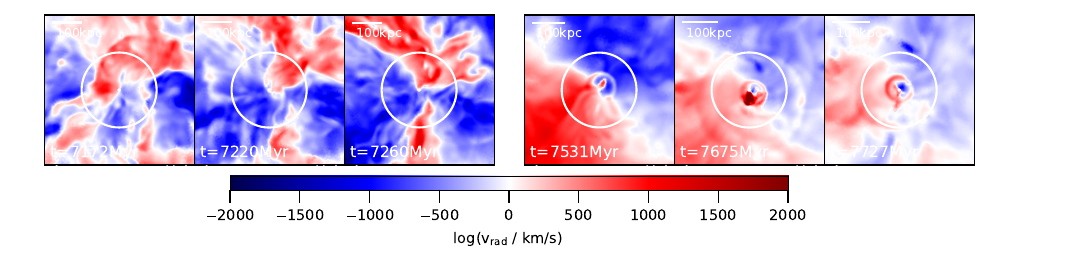}
    \caption{Time evolution of three key physical properties in an example cluster undergoing core entropy transformations, highlighting AGN activity and a merger over a time span of $~550$\,Myr. Time progresses to the left, and covers a time period and the same cluster as depicted in Fig.~\ref{fig:timeseries2}. Each row shows a different property: entropy (top), electron number density (middle), and radial velocity (bottom). All panels are slices of 15\,kpc depth and a radial extent of $0.2\rvir$, with the white circles marking $0.1 \rvir$. The specific (lookback) times shown are selected to highlight significant changes related to AGN feedback -- right three panels -- and mergers -- left three panels of the sequence.}
    \label{fig:VisSeries}
\end{figure*}
\section{Discussion}\label{sec_discussion}

Our analysis shows that, according to TNG-Cluster, clusters may undergo thermodynamical changes in their cores in correspondence to either mergers (Sec.~\ref{sec_mergers}) or AGN feedback activity (Sec.~\ref{sec_agn}). But in which ways should mergers or AGN energy injections have physical effects on the state of the gas at the cores of clusters? And what can we say about the possible non-trivial interplay between mergers and AGN feedback? First, we explore how either mergers or AGN feedback are able to transform a cluster core. We then contextualize our findings from Sec.~\ref{sec_mergers} and Sec.~\ref{sec_agn} with respect to previous results, and finally explore the implications further.

\begin{figure}
    \centering
    \hspace*{-0.5cm}\includegraphics[width=0.46\textwidth]{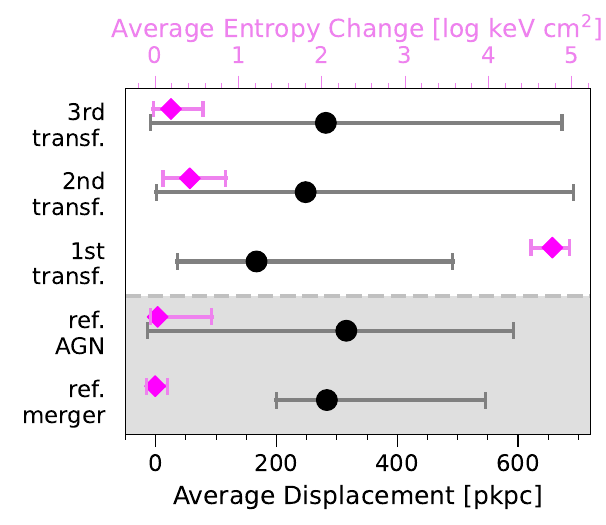}
    \caption{Heating (i.e. entropy increase) versus displacement of the innermost gas during the core transformations of an example cluster, the most massive of TNG300. We first select the central 1,000 gas cells with the lowest entropy. We then measure the difference of the mean halocentric distance (bottom x-axis, black markers) and of the mean entropy (top x-axis, pink) during specific time periods. The first three rows show the three core transformation episodes of the example cluster of Figs.~\ref{fig:timeseries}, \ref{fig:timeseries2} and \ref{fig:VisSeries}. The last two rows provide reference values for the same cluster: one for a period of high AGN activity (fourth row) and another for a phase of strong gas displacement due to a merger (fifth row). The error bars indicate first to third quartile ranges of the distributions at the end of each period of time.}
    \label{fig:DisplacementHeating}
\end{figure}

\subsection{Core gas displacement or heating?}
\label{sec_mechanisms}
What physical processes occur at the center of clusters upon merging or undergoing AGN feedback for them to change their core entropy or even transform from CCs to NCCs? 

In the following we focus on the most frequent core transformations of TNG-Cluster, namely those to higher entropy cores, i.e. in the qualitative direction from CC to NCC states, and we identify two key physical mechanisms: cluster core transformations may be driven (i) by bulk motions moving low-entropy gas out of the core and subsequent mixing of the gas or (ii) by in-situ heating of core gas. It should be noted from the onset that, whereas the former can be easily associated to the case of mergers, energy injections from the central SMBHs may a priori cause both bulk motions i.e. displacements as well as heating.

Visual inspection of the gas evolution throughout the core transformation periods in our simulated systems shows that mergers not only can draw gas into the center and trigger AGN activity, but there are also clear signs of large-scale gas redistribution. In particular, mergers can push cool i.e. low-entropy gas out of the core of a cluster. 
The last rows of Figs.~\ref{fig:timeseries} and \ref{fig:timeseries2} quantify this through the bulk gas velocity within $0.5 r_{\rm 500}$. Particularly in the high-redshift transformations of the most massive cluster of TNG300, we can see pronounced increases in inflow and outflow velocities, indicative of large-scale stirring of the inner intracluster medium.

Fig.~\ref{fig:VisSeries} visualizes the evolution of the same cluster core during these early transformations, showcasing the effects of AGN feedback and the impact of the merger on the core. It shows entropy, electron number density, and radial velocity within the central $0.2\rvir$ across $\sim 1$\,Gyr, with time progressing to the left. In the three columns on the right, clear signs of AGN activity are evident, such as underdense cavities, turbulent stirring, and high-velocity outflows \citep{ayromlou2024,truong2024,prunier2025}. These are time periods where cold, low-entropy, dense gas clouds stream into the cluster center \citep{rohr2024a}, further feeding the SMBH and its feedback activity. In contrast, the three columns on the left visualize the impact of an ongoing merger, where the low-entropy gas from the core is displaced outward as the core structure is disrupted.

In an attempt to disentangle concurring effects, we measure both the large-scale, core-wide displacement and the change of entropy for core gas, during the three core transformations of the most massive cluster of TNG300. Fig.~\ref{fig:DisplacementHeating} selects $\sim 10^{10}\msun$ gas, that is, $\sim 1000$ gas cells closest to the cluster center and with the lowest entropy, and measures their average displacement distance (bottom x-axis) and average entropy change (top x-axis). As a reference, we also consider time periods without transformations but with ongoing mergers (last row) and time periods without transformations but with high AGN activity (fourth row).
In the case of merger activity, there is significant bulk displacement of core gas, but little associated heating. The innermost gas cells are all pushed out of the core with a relatively small scatter in displacements. We argue that in this reference time period, the whole gas is pushed out of the core, because there are no gas cells with small displacements left. On the other hand, during periods of high AGN activity, there is significant variations in displacements of the innermost cluster gas, as only a small fraction of core gas is ejected to larger scales. On the other, the amount and scatter of entropy increase (i.e. heating) is larger in that case.

Focusing on the core transformations of this cluster, we find strong heating during the first transformation, and a wide range of displacements away from the core in all three cases. The diversity of displacements of the innermost gas in such core transformations is at odds with the merging reference case of the last row, whereby merger-induced bulk motions seem to lead to a rather uniform displacement of core gas, with little variation, and also with little heating (last row). On the other hand, as we tend to find significant heating during all three transformation episodes and large ranges of core-gas displacement, we tentatively conclude that, for this cluster and its core transformations, AGN feedback, which in turn could be triggered by mergers, is the direct driver of the changes towards higher-entropy cores.

It is not possible to say how the suggestive and tentative conclusions above -- i.e. that AGN feedback (via heating and partial bulk motions) and not mergers (via core gas displacement) may dominate CCs to NCCs transformations -- generalize to the whole TNG-Cluster population and all their core transformations. Lack of simulation  data, chiefly in the form of high-time cadence output, does not allow to repeat this type of analysis for all simulated clusters we have at hand. Moreover, our proposed analysis remains somewhat qualitative, and so further work is necessary to both understand the proposed physical processes in detail and with what frequency one or the other is the main driver of cluster core transformations.

\subsection{Mergers as drivers of cluster core transformations?}

As mentioned in Sec.~\ref{sec_intro}, mergers have since long been proposed from a numerical and theoretical perspective as possible physical drivers of thermodynamical changes in cluster cores.

With cosmological hydrodynamical simulations of clusters, \citet{burns2007} found that major mergers at early times destroy CCs, while major mergers at late times may not necessarily have the same impact. However, these cosmological simulations have no AGN feedback included, which may make their late time CCs artificially strong against destruction. When AGN feedback is included, late mergers can in general disrupt CCs \citep{rasia2015}. Of the two major mergers studied by \citet{hahn2017}, from two clusters with similar assembly histories, it was concluded that major mergers are a necessary but not sufficient condition to destroy CCs, as only low AM mergers are sufficient. 

Several idealized simulation studies have attempted to identify merger properties that lead to cluster core transformations. \citet{valdarnini2021} found that CCs are destroyed in high-mass major mergers, independent of angular momentum (AM). This is in agreement with our findings, that the mass ratio of mergers is a better predictor for the destruction of a cool core, with AM playing a less important role -- i.e. major mergers more likely destroy a cool core than minor mergers, irrespective of AM. They also concluded that low AM major mergers are the most capable at disrupting CCs. Similarly, we find that major mergers with low AM are more frequently associated to core-disruption transformations. In terms of minor mergers, \citet{valdarnini2021} found that CCs are destroyed in a low AM encounter, but not in a high AM encounter. In contrast, we find that strong transformations in the minor merger regime tend to have high AM, although these also preferentially occur at high halo masses. In TNG-Cluster we find that minor mergers with low AM tend to produce only weak increases in central entropy. Again, however, it is important to note that \citet{valdarnini2021} do not include AGN feedback in their simulations, only stellar feedback.
Finally, \citet{chen2024} find that CCs are: preserved in minor mergers with $1/10$ mass ratios, destroyed in all $1/3$ major mergers, but only destroyed in $1/1$ major mergers with high AM.

The studies and results above were all based on idealized simulations, often without including AGN feedback or without accounting for star formation and stellar feedback. Moreover,they also generally had small sample sizes, by simulating, in most cases, only one merger with a given mass ratio and angular momentum combination. Perhaps partly due to this, they give conflicting results as to the impact of mergers on cluster cores. 

On the other hand, when we consider the 352 halos in TNG-Cluster and their evolutionary histories through billion of years, we find individual examples that reproduce the whole breadth of behaviors seen in all previous simulations. While we find that mergers and transformations are correlated, not every transformation is an event that leads to the disruption of a CC. Furthermore, the properties of the merger that decides whether a CC is disrupted seem more complex than mass ratio and AM alone. 

In terms of cosmological magnetohydrodynamical simulations of large numbers of clusters and their galaxies, \citet{barnes2018} found a similar fraction of relaxed clusters among CCs and NCCs in the TNG300 sample, using various definitions of relaxedness. They argued that this suggests mergers cannot alone be responsible for disrupting CCs. While TNG-Cluster employs the same galaxy formation model as TNG300, we find that mergers can be associated, and hence possibly responsible, for CC disruption in most cases. However, as we cannot associate all transformations with a merger, it remains plausible that not every transformation is merger-driven. In fact, only about 70\% of all transformations in our sample could be linked to a nearby merger. Moreover, we find that mergers temporarily make clusters more unrelaxed, but the halo quickly returns to a relaxed state, at least under the relaxedness definition adopted in this work. This return to relaxedness may occur faster than the time it takes for mergers to fully transform the core. Consequently, NCCs may already appear relaxed even if their cores have not yet returned to CC status, or CCs may appear unrelaxed despite the merger not having had sufficient time to disrupt the core. This rapid decline in the relaxedness measurements suggests that this definition of relaxedness may indeed not be a reliable indicator of CC status.

\subsection{The interplay of AGN feedback and mergers}\label{sec_discAGNMerger}

\citet{chen2024} investigated the combined effects of mergers and AGN feedback in idealized simulations of binary cluster mergers. They identified different scenarios for the transitions between CC and NCC clusters. In all their minor mergers, independent of AM, CCs are preserved. In major mergers with high AM, CCs are destroyed by the combined effect of mergers and AGN. Interestingly, \citet{chen2024} reported cases where low-AM major mergers behave differently: one instance where a low-AM major merger fails to trigger AGN activity yet destroys the CC, and another where a low-AM merger does not generate sufficient heating and thus cannot transform the core. Overall, these diverse scenarios suggest a complex interplay between merger dynamics and AGN feedback.

Our analysis in Sec.~\ref{sec_agn} clearly shows that, within the TNG i.e. TNG-Cluster model, {\it individual} energy injections from the central SMBH of clusters cannot transform cluster cores: energetically, they are not competitive with the total energy required and their impact on central thermodynamical properties, such as $K_0$, is short-lived. However, periods of high SMBH accretion lead to high duty-cycle i.e. rapidly repeating AGN feedback events i.e. frequent subsequent SMBH energy injections. These can indeed produce a collective, long-lasting impact. The combination of many such discrete AGN feedback events within a short time-scale can therefore impact cluster core thermodynamics. This scenario is fully analog, as in fact it is strictly connected, to the case of star formation quenching \citep{nelson2018, terrazas2020, davies2020}, which in TNG is both triggered and maintained by the same pulsated kinetic feedback energy injections \citep{zinger2020}. Also the halting for star formation in TNG massive galaxies is not immediate or instantaneous after a SMBH feedback event: rather, a certain amount of cumulative energy injected by their central SMBH is needed before galaxies are removed of their cold dense gas and hence quenched \citep[see also][]{terrazas2020, piotrowska2022, bluck2023}.

In the previous Sections we have provided quantifications and arguments that support the idea that prolonged AGN activity could transform the core of clusters. However, whether this can happen in the absence of any merger activity is plausible but not demonstrated. For the quenching analogy, for example, it had been shown that mergers are not necessary (and certainly not sufficient) to quench star formation in normal TNG galaxies and only (kinetic) AGN feedback is of the essence \citep{weinberger2017, weinberger2018}\footnote{In fact, only by including a form of large energy injections from the central regions of galaxies it is possible to quench populations of massive galaxies to levels that are consistent with observations.}. However, for the case of the most massive galaxies in the Universe, i.e. for clusters and with TNG-Cluster we cannot demonstrate this: there are always mergers occurring throughout the evolution of clusters and close in time to SMBH energy injections, including the one we can study with fine time cadence within the TNG300 subbox. 

For the particular halo of Figs.~\ref{fig:timeseries} and \ref{fig:timeseries2}, there is always a merger that also occurs somewhat prior to each transformation towards higher core entropy. Disentangling the effects of mergers and AGN activity is therefore difficult, especially since they are coupled to each other. For this halo, there are several possible scenarios for how a merger can transform the core: (i) Mergers could trigger core transformations alone, or (ii) they could trigger high AGN activity \textit{and} core transformations at the same time, or (iii) merger-increased AGN activity could aid in core transformation due to the merger, while being unable to do so alone. (iv) Likewise, mergers may be unable to disrupt cool-cores \textit{unless} the corresponding AGN activity is high, and simultaneously `softening' the core.

As we have shown in Section~\ref{sec_mechanisms} and Fig.~\ref{fig:DisplacementHeating}, this particular halo has three transformations to higher core entropy that are tentatively causally linked to AGN activity. In fact, that particular halo has many more numerous mergers than three. That is, some (major) mergers are not associated with core transformations. In fact, the single merger with the largest mass ratio and lowest angular momentum is unassociated with any of the three core transformations, and occurs during a period of low AGN activity. In contrast, the three mergers we associate with the three transformations of this cluster have higher AM. Possibly, these events pull gas into the core of the cluster, thereby inducing AGN activity. This suggests that mergers are, in this case, a necessary but not sufficient condition for core transformation, as is AGN feedback. Rather, both actively work together to transform the cluster core -- at least in this case.

Putting together all our findings, we tentatively suggest (at least) two  physical mechanisms that can successfully transform a cluster from CC to NCC. First, (i) mergers generate large-scale bulk motions that displace low-entropy gas from the cluster center. These events result in moderate heating through mixing but do not significantly activate AGN feedback. The merger is the main driver of the transformation in this case. This scenario may possibly found more often among major mergers (with low angular momentum). In contrast, (ii) extended periods of AGN activity substantially raise the entropy of the central gas, leading eventually to the destruction of the cool core. In this scenario, mergers and AGN feedback may work together, in that it is possible that episodes of high duty-cycle AGN activity are triggered by mergers (chiefly high angular momentum (minor) mergers) but not exclusively i.e. mergers may not be necessary in all instances to drive gas towards cluster cores (smooth gas-accretion flows from the large-scale structure might in principle do the job).

These are two tentative scenarios, but the results of this work \citep[and of others, e.g.][]{sanchez2021} suggest a possible greater diversity of processes capable of destroying CCs. Regrettably, despite the large statistics, physics completeness, and realism outcome of TNG-Cluster, we cannot ultimately and quantitatively say whether cluster CCs are more frequently destroyed via mergers and gas displacement or via SMBH feedback, in turn triggered or not by mergers. We suspect that AGN feedback (via heating and partial bulk motions) dominates CCs to NCCs transformations, at least in TNG-Cluster and as we could show in the case of a couple of example clusters. However, we cannot ultimately demonstrate and quantify it.


\section{Summary and conclusions} \label{sec_conclusion}

In this paper we study the thermodynamical transformations of the cores of 352 galaxy clusters from the new TNG-Cluster cosmological magnetohydrodynamical simulation of galaxies. We identify core transformations based on rapid changes in central entropy (Fig.~\ref{fig:K0vsTime}) as clusters evolve from low entropy (e.g. cool-cores, CCs) to high entropy (e.g. non-cool-cores, NCCs) states and vise versa. We study the statistics of these transformations across our large sample of simulated galaxy clusters, in the $\mvir = 10^{14-15.3}~\msun$ mass range and since the simulated clusters passed the mass threshold of $10^{14}~\msun$, i.e. our cluster formation redshift: $z\lesssim2.5$ for all studied clusters. We then investigate time correlations with merger and AGN feedback activity. Our main findings are:

\begin{itemize}
    \item \textbf{The overall TNG-Cluster population evolves to higher core entropy.} We demonstrate that the overall population of galaxy clusters in TNG-Cluster evolves to higher core entropy as time progresses. This is consistent with a population of halos that are mainly born as CC clusters and evolves towards the NCC regime over time (Figs.~\ref{fig:M500VsZform} and \ref{fig:K0vsTimeAllHalos}). 
    \item \textbf{Clusters frequently undergo transformations.} In our sample of 352 clusters, we identify a total of 478 transformations to higher core entropy, i.e. in the direction from CC to NCC states, and 97 transformations towards lower core entropy, in the direction from NCC towards CC states. On average, each cluster undergoes 2 to 3 transformations during its lifetime, and more massive clusters experience more transformations (Figs.~\ref{fig:ChangeRate} and ~\ref{fig:M500VsNumTrafo}). 
    \item \textbf{Transformations have diverse timescales, and can be temporary.} We find a broad range of transformation durations, from changes as fast as $\sim 0.5$\,Gyr up to $\sim 4$\,Gyr (Fig.~\ref{fig:TimescaleChange}). Slower transformation episodes tend to occur at lower redshift. After a core transformation, clusters can occasionally revert and return to their earlier CC or NCC state. Temporary CC phases are shorter ($\sim 1-2$\,Gyr) than temporary NCC phases ($\sim 1-8$\,Gyr, Fig.~\ref{fig:HistDuration}).
    \item \textbf{The evolution of cluster core thermodynamics is diverse.} We identify six different archetypes of clusters based on their evolutionary histories (Table~\ref{tab:archetypes}). The most common class ($64\%$) are clusters that undergo only transformations to higher core entropy (i.e. towards NCC states), whereas the least common type ($4\%$) transitions only to lower core entropy (i.e. towards CC states). We find in our sample $10\%$ clusters that do not transform at all. Another $10\%$ frequently undergo transformation in both possible directions. The cluster classes with the most transformations are also the clusters with the most mergers (Fig.~\ref{fig:K0vsM500vsTrafoClasses}).
    \item \textbf{Mergers and transformations are connected.} Changes to higher core entropy are closely linked to mergers. In our sample, pairs of mergers and transitions to higher core entropy occurring within less than 1\,Gyr are more frequent than would be expected from a random sample. This is particularly true for major mergers (Figs.~\ref{fig:fracTwwoM} and \ref{fig:cumHistdtMerger}).
    \item \textbf{Diversity in mergers that transform cores.} We find indications and trends that major mergers are more effective in destroying cool cores than minor mergers and that major mergers with low angular momentum are somewhat the most effective. However, most importantly, we find examples of mergers spanning \textit{all} mass ratios and angular momenta that can be associated, and possibly successfully produce, core transformations, including CC disruption (Fig.~\ref{fig:HistMR}).
    \item \textbf{AGN feedback as a driver of core transformations.} We study the time evolution of a prototypical cluster, where we can track AGN feedback in detail and with temporal cadence of 5-10 Myr. Individual AGN energy injection events cause short-lived increases in central entropy (Figs.~\ref{fig:timeseries} and \ref{fig:timeseries2}). However, the collective effect of numerous AGN feedback episodes during extended periods of efficient accretion leads to a short duty-cycle between successive energy injections, enabling AGN feedback to produce long-term enhancement of core entropy (Figs.~\ref{fig:timeseries}, \ref{fig:timeseries2}, \ref{fig:TimeDiffAGNBursts}). 
    \item \textbf{Physical processes responsible for enhancing core entropy.} We tentatively show that merger-induced bulk motions can displace core gas, hence removing low-entropy gas from the cluster centers with limited heating. In contrast, AGN feedback produces significant heating in addition to fast outflows. This results in a broader range of displacements of the innermost gas (Figs.~\ref{fig:VisSeries} and \ref{fig:DisplacementHeating}). Studying our one prototypical cluster in detail, we find significant heating and large ranges of displacement during all core transformation episodes. This suggests that AGN feedback directly drives the changes towards higher-entropy cores, whether triggered by a preceding merger or not, at least for this particular cluster.
   
\end{itemize}

Our previous analysis of CC versus NCC statistics in TNG-Cluster indicates that CCs and NCCs are not two distinct classes \citep{lehle2024}. This work demonstrates how clusters can indeed frequently change their core states, in either direction, with diverse timescales, and more than once across their history. Overall, however, core transformations towards high entropy are more frequent and long lasting, consistently with a TNG-Cluster population that moves towards NCC states at lower redshifts. 

The majority (70\%) but not all the identified core transformations in TNG-Cluster can be linked, time-wise, to a merger: thanks to the large statistics of TNG-Cluster, we can find examples of the whole breadth of behaviors, i.e. merger configurations versus core entropy changes, seen in all previous simulations, but still with major mergers being more frequently associated with the destruction of CCs. However, the number of mergers a cluster experienced is not a good  predictor for $z=0$ CC status, i.e. it is only in the case of many mergers above the average, in which case clusters end up at $z=0$ as NCCs.

Despite the richness and realism of the TNG-Cluster outcome, we cannot ultimately pin down whether CCs are more frequently destroyed via mergers and gas displacement or via SMBH feedback. We tentatively propose that AGN feedback (via heating and partial bulk motions), in turn triggered or not by mergers, may be the dominant and necessary driver for CCs to NCCs transformations across the cluster population, but we can show this only for a few core changes, and somewhat qualitatively. 

In addition to cool-core disruption, the physical drivers of core transformations in the opposite direction (from high to low entropy) remain an open question, i.e. how clusters can {\it rejuvenate} their cool-cores. Could a cluster always return to a cool-core state, given sufficient time and in the absence of energy injections? In this work we have shown that the timescales distribution of the transformations towards low entropy is bimodal -- could this imply two distinct types of events, with different drivers? 

Future work with TNG-Cluster and new simulations extending the TNG-Cluster physics may help unraveling all outstanding details of the complex and dynamic life of galaxy clusters, and their cores, that we have revealed in this paper. 

\section*{Data Availability}

The IllustrisTNG simulations themselves are publicly available and accessible at \url{www.tng-project.org/data} since a few years \citep{nelson2019a}: the TNG-Cluster simulation and the data directly related to this work is also public and accessible there.

\section*{Acknowledgements}

KL is a Fellow of the International Max Planck
Research School for Astronomy and Cosmic Physics at the University of Heidelberg (IMPRS-HD) and acknowledges funding from the Hector Fellow Academy through a Research Career Development Award. DN acknowledges funding from the Deutsche Forschungsgemeinschaft (DFG) through an Emmy Noether Research Group (grant number NE 2441/1-1). This work is also supported by the European Union (ERC, COSMIC-KEY, 101087822, PI: Pillepich).

The TNG-Cluster simulation has been run on several computer clusters: as part of the TNG-Cluster project on the HoreKa supercomputer, funded by the Ministry of Science, Research and the Arts Baden-Württemberg and by the Federal Ministry of Education and Research. The bwForCluster Helix supercomputer, supported by the state of Baden-Württemberg through bwHPC and the German Research Foundation (DFG) through grant INST 35/1597-1~FUGG. The Vera, Cobra, and Raven clusters of the Max Planck Computational Data Facility (MPCDF). The BinAC cluster, supported by the High Performance and Cloud Computing Group at the Zentrum für Datenverarbeitung of the University of Tübingen, the state of Baden-Württemberg through bwHPC and the German Research Foundation (DFG) through grant no INST 37/935-1~FUGG. This analysis has been carried out on the Vera supercomputer of the Max Planck Institute for Astronomy (MPIA).

\bibliographystyle{aa}
\bibliography{refs}

\begin{thebibliography}{74}
\expandafter\ifx\csname natexlab\endcsname\relax\def\natexlab#1{#1}\fi

\bibitem[{Allen {et~al.}(2001)Allen, Fabian, Johnstone, Arnaud, \& Nulsen}]{allen2001}
Allen, S.~W., Fabian, A.~C., Johnstone, R.~M., Arnaud, K.~A., \& Nulsen, P. E.~J. 2001, \href{http://dx.doi.org/10.1046/j.1365-8711.2001.04135.x}{\color{blue}MNRAS}, 322, 589

\bibitem[{{Andrade-Santos} {et~al.}(2017){Andrade-Santos}, Jones, Forman, Lovisari, Vikhlinin, Weeren, Murray, Arnaud, Pratt, D{\'e}mocl{\`e}s, Kraft, Mazzotta, B{\"o}hringer, Chon, Giacintucci, Clarke, Borgani, David, Douspis, Pointecouteau, Dahle, Brown, Aghanim, \& Rasia}]{andrade-santos2017}
{Andrade-Santos}, F., Jones, C., Forman, W.~R., {et~al.} 2017, \href{http://dx.doi.org/10.3847/1538-4357/aa7461}{\color{blue}ApJ}, 843, 76

\bibitem[{Ayromlou {et~al.}(2024)Ayromlou, Nelson, Pillepich, Rohr, Truong, Li, Simionescu, Lehle, \& Lee}]{ayromlou2024}
Ayromlou, M., Nelson, D., Pillepich, A., {et~al.} 2024, \href{http://dx.doi.org/10.1051/0004-6361/202348612}{\color{blue}A\&A}, 690, A20

\bibitem[{Barai {et~al.}(2016)Barai, Murante, Borgani, Gaspari, Granato, Monaco, \& {Ragone-Figueroa}}]{barai2016}
Barai, P., Murante, G., Borgani, S., {et~al.} 2016, \href{http://dx.doi.org/10.1093/mnras/stw1389}{\color{blue}MNRAS}, 461, 1548

\bibitem[{Barnes {et~al.}(2017)Barnes, Kay, Bah{\'e}, Dalla~Vecchia, McCarthy, Schaye, Bower, Jenkins, Thomas, Schaller, Crain, Theuns, \& White}]{barnes2017a}
Barnes, D.~J., Kay, S.~T., Bah{\'e}, Y.~M., {et~al.} 2017, \href{http://dx.doi.org/10.1093/mnras/stx1647}{\color{blue}MNRAS}, 471, 1088

\bibitem[{Barnes {et~al.}(2018)Barnes, Vogelsberger, Kannan, Marinacci, Weinberger, Springel, Torrey, Pillepich, Nelson, Pakmor, Naiman, Hernquist, \& McDonald}]{barnes2018}
Barnes, D.~J., Vogelsberger, M., Kannan, R., {et~al.} 2018, \href{http://dx.doi.org/10.1093/mnras/sty2078}{\color{blue}MNRAS}, 481, 1809

\bibitem[{Bartalucci {et~al.}(2023)Bartalucci, Molendi, Rasia, Pratt, Arnaud, Rossetti, Gastaldello, Eckert, Balboni, Borgani, Bourdin, Campitiello, De~Grandi, De~Petris, Duffy, Ettori, Ferragamo, Gaspari, Gavazzi, Ghizzardi, Iqbal, Kay, Lovisari, Mazzotta, Maughan, Pointecouteau, Riva, \& Sereno}]{bartalucci2023}
Bartalucci, I., Molendi, S., Rasia, E., {et~al.} 2023, {{CHEX-MATE}}: {{Constraining}} the Origin of the Scatter in Galaxy Cluster Radial {{X-ray}} Surface Brightness Profiles

\bibitem[{Bluck {et~al.}(2023)Bluck, Piotrowska, \& Maiolino}]{bluck2023}
Bluck, A. F.~L., Piotrowska, J.~M., \& Maiolino, R. 2023, \href{http://dx.doi.org/10.3847/1538-4357/acac7c}{\color{blue}Astrophys. J.}, 944, 108

\bibitem[{Burns {et~al.}(2007)Burns, Hallman, Gantner, Motl, \& Norman}]{burns2007}
Burns, J., Hallman, E., Gantner, B., Motl, P., \& Norman, M. 2007, \href{http://dx.doi.org/10.1086/526514}{\color{blue}The Astrophysical Journal}, 675

\bibitem[{Chadayammuri {et~al.}(2021)Chadayammuri, Tremmel, Nagai, Babul, \& Quinn}]{chadayammuri2021}
Chadayammuri, U., Tremmel, M., Nagai, D., Babul, A., \& Quinn, T. 2021, \href{http://dx.doi.org/10.1093/mnras/stab1010}{\color{blue}MNRAS}, 504, 3922

\bibitem[{Chen {et~al.}(2024)Chen, Yang, Schive, ZuHone, \& Gaspari}]{chen2024}
Chen, S.-S., Yang, H.-Y.~K., Schive, H.-Y., ZuHone, J., \& Gaspari, M. 2024, Cool-{{Core Destruction}} in {{Merging Clusters}} with {{AGN Feedback}} and {{Radiative Cooling}}

\bibitem[{Cowie \& Binney(1977)}]{cowie1977}
Cowie, L.~L. \& Binney, J. 1977, \href{http://dx.doi.org/10.1086/155406}{\color{blue}Astrophys. J.}, 215, 723

\bibitem[{Davies {et~al.}(2020)Davies, Crain, Oppenheimer, \& Schaye}]{davies2020}
Davies, J.~J., Crain, R.~A., Oppenheimer, B.~D., \& Schaye, J. 2020, \href{http://dx.doi.org/10.1093/mnras/stz3201}{\color{blue}MNRAS}, 491, 4462

\bibitem[{Eckert {et~al.}(2011)Eckert, Molendi, \& Paltani}]{eckert2011}
Eckert, D., Molendi, S., \& Paltani, S. 2011, \href{http://dx.doi.org/10.1051/0004-6361/201015856}{\color{blue}A\&A}, 526, A79

\bibitem[{Ehlert {et~al.}(2023)Ehlert, Weinberger, Pfrommer, Pakmor, \& Springel}]{ehlert2023}
Ehlert, K., Weinberger, R., Pfrommer, C., Pakmor, R., \& Springel, V. 2023, \href{http://dx.doi.org/10.1093/mnras/stac2860}{\color{blue}MNRAS}, 518, 4622

\bibitem[{Fabian \& Nulsen(1977)}]{fabian1977}
Fabian, A.~C. \& Nulsen, P. E.~J. 1977, \href{http://dx.doi.org/10.1093/mnras/180.3.479}{\color{blue}MNRAS}, 180, 479

\bibitem[{Graham {et~al.}(2023)Graham, O'Donnell, Silverstein, Eiger, Jeltema, Hollowood, Cross, Everett, Giles, Jobel, Laubner, McDaniel, Romer, Swart, Aguena, Allam, Alves, Brooks, Kind, Carretero, Costanzi, {da Costa}, Pereira, De~Vicente, Desai, Dietrich, Doel, Ferrero, Frieman, {Garcia-Bellido}, Gruen, Gruendl, Hinton, Honscheid, James, Kuehn, Kuropatkin, Lahav, Marshall, Melchior, {Mena-Fernandez}, Menanteau, Miquel, Ogando, Palmese, Pieres, Malagon, Reil, {Rodriguez-Monroy}, Sanchez, Scarpine, Schubnell, Smith, Suchyta, Tarle, To, \& Weaverdyck}]{graham2023}
Graham, K., O'Donnell, J., Silverstein, M.~M., {et~al.} 2023, Cool {{Cores}} in {{Clusters}} of {{Galaxies}} in the {{Dark Energy Survey}}

\bibitem[{Guo \& Mathews(2010)}]{guo2010}
Guo, F. \& Mathews, W.~G. 2010, \href{http://dx.doi.org/10.1088/0004-637X/717/2/937}{\color{blue}ApJ}, 717, 937

\bibitem[{Guo \& Oh(2009)}]{guo2009}
Guo, F. \& Oh, S.~P. 2009, \href{http://dx.doi.org/10.1111/j.1365-2966.2009.15592.x}{\color{blue}MNRAS}, 400, 1992

\bibitem[{Hahn {et~al.}(2017)Hahn, Martizzi, Wu, Evrard, Teyssier, \& Wechsler}]{hahn2017}
Hahn, O., Martizzi, D., Wu, H.-Y., {et~al.} 2017, \href{http://dx.doi.org/10.1093/mnras/stx001}{\color{blue}MNRAS}, stx001

\bibitem[{Hudson {et~al.}(2010)Hudson, Mittal, Reiprich, Nulsen, Andernach, \& Sarazin}]{hudson2010}
Hudson, D.~S., Mittal, R., Reiprich, T.~H., {et~al.} 2010, \href{http://dx.doi.org/10.1051/0004-6361/200912377}{\color{blue}A\&A}, 513, A37

\bibitem[{Hudson \& Reiprich(2007)}]{hudson2007}
Hudson, D.~S. \& Reiprich, T.~H. 2007, Investigating the {{Central Regions}} of the {{HIFLUGCS Clusters}} with {{Chandra}}, 42

\bibitem[{Lea {et~al.}(1973)Lea, Silk, Kellogg, \& Murray}]{lea1973}
Lea, S.~M., Silk, J., Kellogg, E., \& Murray, S. 1973, \href{http://dx.doi.org/10.1086/181300}{\color{blue}Astrophys. J.}, 184, L105

\bibitem[{Lee {et~al.}(2024)Lee, Pillepich, ZuHone, Nelson, Jee, Nagai, \& Finner}]{lee2024}
Lee, W., Pillepich, A., ZuHone, J., {et~al.} 2024, \href{http://dx.doi.org/10.1051/0004-6361/202348194}{\color{blue}A\&A}, 686, A55

\bibitem[{Lehle {et~al.}(2024)Lehle, Nelson, Pillepich, Truong, \& Rohr}]{lehle2024}
Lehle, K., Nelson, D., Pillepich, A., Truong, N., \& Rohr, E. 2024, \href{http://dx.doi.org/10.1051/0004-6361/202348609}{\color{blue}A\&A}, 687, A129

\bibitem[{Lin {et~al.}(2015)Lin, McDonald, Benson, \& Miller}]{lin2015}
Lin, H.~W., McDonald, M., Benson, B., \& Miller, E. 2015, \href{http://dx.doi.org/10.1088/0004-637X/802/1/34}{\color{blue}ApJ}, 802, 34

\bibitem[{Liu {et~al.}(2024)Liu, Sun, Voit, Lal, Nulsen, Gaspari, Sarazin, Ehlert, \& Zheng}]{liu2024}
Liu, W., Sun, M., Voit, G.~M., {et~al.} 2024, X-Ray {{Cool Core Remnants Heated}} by {{Strong Radio AGN Feedback}}

\bibitem[{Marinacci {et~al.}(2018)Marinacci, Vogelsberger, Pakmor, Torrey, Springel, Hernquist, Nelson, Weinberger, Pillepich, Naiman, \& Genel}]{marinacci2018}
Marinacci, F., Vogelsberger, M., Pakmor, R., {et~al.} 2018, \href{http://dx.doi.org/10.1093/mnras/sty2206}{\color{blue}MNRAS}

\bibitem[{McCarthy {et~al.}(2004)McCarthy, Balogh, Babul, Poole, \& Horner}]{mccarthy2004}
McCarthy, I.~G., Balogh, M.~L., Babul, A., Poole, G.~B., \& Horner, D.~J. 2004, \href{http://dx.doi.org/10.1086/423267}{\color{blue}ApJ}, 613, 811

\bibitem[{McDonald {et~al.}(2017)McDonald, Allen, Bayliss, Benson, Bleem, Brodwin, Bulbul, Carlstrom, Forman, {Hlavacek-Larrondo}, Garmire, Gaspari, Gladders, Mantz, \& Murray}]{mcdonald2017}
McDonald, M., Allen, S.~W., Bayliss, M., {et~al.} 2017, \href{http://dx.doi.org/10.3847/1538-4357/aa7740}{\color{blue}ApJ}, 843, 28

\bibitem[{McDonald {et~al.}(2013)McDonald, Benson, Vikhlinin, Stalder, Bleem, Lin, Aird, Ashby, Bautz, Bayliss, Bocquet, Brodwin, Carlstrom, Chang, Cho, Clocchiatti, Crawford, Crites, {de Haan}, Desai, Dobbs, Dudley, Foley, Forman, George, Gettings, Gladders, Gonzalez, Halverson, High, Holder, Holzapfel, Hoover, Hrubes, Jones, Joy, Keisler, Knox, Lee, Leitch, Liu, Lueker, {Luong-Van}, Mantz, Marrone, McMahon, Mehl, Meyer, Miller, Mocanu, Mohr, Montroy, Murray, Nurgaliev, Padin, Plagge, Pryke, Reichardt, Rest, Ruel, Ruhl, Saliwanchik, Saro, Sayre, Schaffer, Shirokoff, Song, Suhada, Spieler, Stanford, Staniszewski, Stark, Story, {van Engelen}, Vanderlinde, Vieira, Williamson, Zahn, \& Zenteno}]{mcdonald2013}
McDonald, M., Benson, B.~A., Vikhlinin, A., {et~al.} 2013, \href{http://dx.doi.org/10.1088/0004-637X/774/1/23}{\color{blue}ApJ}, 774, 23

\bibitem[{McNamara {et~al.}(2009)McNamara, Kazemzadeh, Rafferty, B{\^i}rzan, Nulsen, Kirkpatrick, \& Wise}]{mcnamara2009}
McNamara, B.~R., Kazemzadeh, F., Rafferty, D.~A., {et~al.} 2009, \href{http://dx.doi.org/10.1088/0004-637X/698/1/594}{\color{blue}ApJ}, 698, 594

\bibitem[{Mohr {et~al.}(1999)Mohr, Mathiesen, \& Evrard}]{mohr1999}
Mohr, J.~J., Mathiesen, B., \& Evrard, A.~E. 1999, \href{http://dx.doi.org/10.1086/307227}{\color{blue}Astrophys. J.}, 517, 627

\bibitem[{Molendi {et~al.}(2022)Molendi, De~Grandi, Rossetti, Bartalucci, Gastaldello, Ghizzardi, \& Gaspari}]{molendi2022}
Molendi, S., De~Grandi, S., Rossetti, M., {et~al.} 2022, The Evolving Cluster Cores: {{Putting}} Together the Pieces of the Puzzle

\bibitem[{Molendi \& Pizzolato(2001)}]{molendi2001}
Molendi, S. \& Pizzolato, F. 2001, \href{http://dx.doi.org/10.1086/322387}{\color{blue}ApJ}, 560, 194

\bibitem[{Mushotzky {et~al.}(1978)Mushotzky, Serlemitsos, Smith, Boldt, \& Holt}]{mushotzky1978}
Mushotzky, R.~F., Serlemitsos, P.~J., Smith, B.~W., Boldt, E.~A., \& Holt, S.~S. 1978, \href{http://dx.doi.org/10.1086/156465}{\color{blue}Astrophys. J.}, 225, 21

\bibitem[{Naiman {et~al.}(2018)Naiman, Pillepich, Springel, {Ramirez-Ruiz}, Torrey, Vogelsberger, Pakmor, Nelson, Marinacci, Hernquist, Weinberger, \& Genel}]{naiman2018}
Naiman, J.~P., Pillepich, A., Springel, V., {et~al.} 2018, \href{http://dx.doi.org/10.1093/mnras/sty618}{\color{blue}MNRAS}, 477, 1206

\bibitem[{Nelson {et~al.}(2024)Nelson, Pillepich, Ayromlou, Lee, Lehle, Rohr, \& Truong}]{nelson2024}
Nelson, D., Pillepich, A., Ayromlou, M., {et~al.} 2024, \href{http://dx.doi.org/10.1051/0004-6361/202348608}{\color{blue}A\&A}, 686, A157

\bibitem[{Nelson {et~al.}(2019{\natexlab{a}})Nelson, Pillepich, Springel, Pakmor, Weinberger, Genel, Torrey, Vogelsberger, Marinacci, \& Hernquist}]{nelson2019b}
Nelson, D., Pillepich, A., Springel, V., {et~al.} 2019{\natexlab{a}}, \href{http://dx.doi.org/10.1093/mnras/stz2306}{\color{blue}MNRAS}, 490, 3234

\bibitem[{Nelson {et~al.}(2019{\natexlab{b}})Nelson, Pillepich, Springel, Pakmor, Weinberger, Genel, Torrey, Vogelsberger, Marinacci, \& Hernquist}]{nelson2019}
Nelson, D., Pillepich, A., Springel, V., {et~al.} 2019{\natexlab{b}}, \href{http://dx.doi.org/10.1093/mnras/stz2306}{\color{blue}MNRAS}, 490, 3234

\bibitem[{Nelson {et~al.}(2018)Nelson, Pillepich, Springel, Weinberger, Hernquist, Pakmor, Genel, Torrey, Vogelsberger, Kauffmann, Marinacci, \& Naiman}]{nelson2018}
Nelson, D., Pillepich, A., Springel, V., {et~al.} 2018, \href{http://dx.doi.org/10.1093/mnras/stx3040}{\color{blue}MNRAS}, 475, 624

\bibitem[{Nelson {et~al.}(2019{\natexlab{c}})Nelson, Springel, Pillepich, {Rodriguez-Gomez}, Torrey, Genel, Vogelsberger, Pakmor, Marinacci, Weinberger, Kelley, Lovell, Diemer, \& Hernquist}]{nelson2019a}
Nelson, D., Springel, V., Pillepich, A., {et~al.} 2019{\natexlab{c}}, The {{IllustrisTNG Simulations}}: {{Public Data Release}}

\bibitem[{Nulsen {et~al.}(2005)Nulsen, McNamara, Wise, \& David}]{nulsen2005}
Nulsen, P. E.~J., McNamara, B.~R., Wise, M.~W., \& David, L.~P. 2005, \href{http://dx.doi.org/10.1086/430845}{\color{blue}ApJ}, 628, 629

\bibitem[{Pakmor {et~al.}(2011)Pakmor, Bauer, \& Springel}]{pakmor2011}
Pakmor, R., Bauer, A., \& Springel, V. 2011, \href{http://dx.doi.org/10.1111/j.1365-2966.2011.19591.x}{\color{blue}MNRAS}, 418, 1392

\bibitem[{Pakmor \& Springel(2013)}]{pakmor2013}
Pakmor, R. \& Springel, V. 2013, \href{http://dx.doi.org/10.1093/mnras/stt428}{\color{blue}MNRAS}, 432, 176

\bibitem[{Pascut \& Ponman(2015)}]{pascut2015}
Pascut, A. \& Ponman, T.~J. 2015, \href{http://dx.doi.org/10.1093/mnras/stu2688}{\color{blue}MNRAS}, 447, 3723

\bibitem[{Pillepich {et~al.}(2018{\natexlab{a}})Pillepich, Nelson, Hernquist, Springel, Pakmor, Torrey, Weinberger, Genel, Naiman, Marinacci, \& Vogelsberger}]{pillepich2018a}
Pillepich, A., Nelson, D., Hernquist, L., {et~al.} 2018{\natexlab{a}}, \href{http://dx.doi.org/10.1093/mnras/stx3112}{\color{blue}MNRAS}, 475, 648

\bibitem[{Pillepich {et~al.}(2019)Pillepich, Nelson, Springel, Pakmor, Torrey, Weinberger, Vogelsberger, Marinacci, Genel, {van~der~Wel}, \& Hernquist}]{pillepich2019}
Pillepich, A., Nelson, D., Springel, V., {et~al.} 2019, \href{http://dx.doi.org/10.1093/mnras/stz2338}{\color{blue}MNRAS}, 490, 3196

\bibitem[{Pillepich {et~al.}(2021)Pillepich, Nelson, Truong, Weinberger, {Martin-Navarro}, Springel, Faber, \& Hernquist}]{pillepich2021}
Pillepich, A., Nelson, D., Truong, N., {et~al.} 2021, \href{http://dx.doi.org/10.1093/mnras/stab2779}{\color{blue}MNRAS}, 508, 4667

\bibitem[{Pillepich {et~al.}(2018{\natexlab{b}})Pillepich, Springel, Nelson, Genel, Naiman, Pakmor, Hernquist, Torrey, Vogelsberger, Weinberger, \& Marinacci}]{pillepich2018}
Pillepich, A., Springel, V., Nelson, D., {et~al.} 2018{\natexlab{b}}, \href{http://dx.doi.org/10.1093/mnras/stx2656}{\color{blue}MNRAS}, 473, 4077

\bibitem[{Piotrowska {et~al.}(2022)Piotrowska, Bluck, Maiolino, \& Peng}]{piotrowska2022}
Piotrowska, J.~M., Bluck, A. F.~L., Maiolino, R., \& Peng, Y. 2022, \href{http://dx.doi.org/10.1093/mnras/stab3673}{\color{blue}MNRAS}, 512, 1052

\bibitem[{{Planck Collaboration} {et~al.}(2016){Planck Collaboration}, Ade, Aghanim, Arnaud, Ashdown, Aumont, Baccigalupi, Banday, Barreiro, Bartlett, Bartolo, Battaner, Battye, Benabed, Beno{\^i}t, {Benoit-L{\'e}vy}, Bernard, Bersanelli, Bielewicz, Bock, Bonaldi, Bonavera, Bond, Borrill, Bouchet, Boulanger, Bucher, Burigana, Butler, Calabrese, Cardoso, Catalano, Challinor, Chamballu, Chary, Chiang, Chluba, Christensen, Church, Clements, Colombi, Colombo, Combet, Coulais, Crill, Curto, Cuttaia, Danese, Davies, Davis, De~Bernardis, De~Rosa, De~Zotti, Delabrouille, D{\'e}sert, Di~Valentino, Dickinson, Diego, Dolag, Dole, Donzelli, Dor{\'e}, Douspis, Ducout, Dunkley, Dupac, Efstathiou, Elsner, En{\ss}lin, Eriksen, Farhang, Fergusson, Finelli, Forni, Frailis, Fraisse, Franceschi, Frejsel, Galeotta, Galli, Ganga, Gauthier, Gerbino, Ghosh, Giard, {Giraud-H{\'e}raud}, Giusarma, Gjerl{\o}w, {Gonz{\'a}lez-Nuevo}, G{\'o}rski, Gratton, Gregorio, Gruppuso, Gudmundsson, Hamann, Hansen, Hanson, Harrison, Helou,
  {Henrot-Versill{\'e}}, {Hern{\'a}ndez-Monteagudo}, Herranz, Hildebrandt, Hivon, Hobson, Holmes, Hornstrup, Hovest, Huang, Huffenberger, Hurier, Jaffe, Jaffe, Jones, Juvela, Keih{\"a}nen, Keskitalo, Kisner, Kneissl, Knoche, Knox, Kunz, {Kurki-Suonio}, Lagache, L{\"a}hteenm{\"a}ki, Lamarre, Lasenby, Lattanzi, Lawrence, Leahy, Leonardi, Lesgourgues, Levrier, Lewis, Liguori, Lilje, {Linden-V{\o}rnle}, {L{\'o}pez-Caniego}, Lubin, {Mac{\'i}as-P{\'e}rez}, Maggio, Maino, Mandolesi, Mangilli, Marchini, Maris, Martin, Martinelli, {Mart{\'i}nez-Gonz{\'a}lez}, Masi, Matarrese, McGehee, Meinhold, Melchiorri, Melin, Mendes, Mennella, Migliaccio, Millea, Mitra, {Miville-Desch{\^e}nes}, Moneti, Montier, Morgante, Mortlock, Moss, Munshi, Murphy, Naselsky, Nati, Natoli, Netterfield, {N{\o}rgaard-Nielsen}, Noviello, Novikov, Novikov, Oxborrow, Paci, Pagano, Pajot, Paladini, Paoletti, Partridge, Pasian, Patanchon, Pearson, Perdereau, Perotto, Perrotta, Pettorino, Piacentini, Piat, Pierpaoli, Pietrobon, Plaszczynski,
  Pointecouteau, Polenta, Popa, Pratt, Pr{\'e}zeau, Prunet, Puget, Rachen, Reach, Rebolo, Reinecke, Remazeilles, Renault, Renzi, Ristorcelli, Rocha, Rosset, Rossetti, Roudier, {Rouill{\'e} d'Orfeuil}, {Rowan-Robinson}, {Rubi{\~n}o-Mart{\'i}n}, Rusholme, Said, Salvatelli, Salvati, Sandri, Santos, Savelainen, Savini, Scott, Seiffert, Serra, Shellard, Spencer, Spinelli, Stolyarov, Stompor, Sudiwala, Sunyaev, Sutton, {Suur-Uski}, Sygnet, Tauber, Terenzi, Toffolatti, Tomasi, Tristram, Trombetti, Tucci, Tuovinen, T{\"u}rler, Umana, Valenziano, Valiviita, Van~Tent, Vielva, Villa, Wade, Wandelt, Wehus, White, White, Wilkinson, Yvon, Zacchei, \& Zonca}]{planckcollaboration2016}
{Planck Collaboration}, Ade, P. A.~R., Aghanim, N., {et~al.} 2016, \href{http://dx.doi.org/10.1051/0004-6361/201525830}{\color{blue}A\&A}, 594, A13

\bibitem[{Poole {et~al.}(2008)Poole, Babul, McCarthy, Sanderson, \& Fardal}]{poole2008}
Poole, G.~B., Babul, A., McCarthy, I.~G., Sanderson, A. J.~R., \& Fardal, M.~A. 2008, \href{http://dx.doi.org/10.1111/j.1365-2966.2008.14003.x}{\color{blue}MNRAS}, 391, 1163

\bibitem[{Prunier {et~al.}(2025)Prunier, {Hlavacek-Larrondo}, Pillepich, Lehle, \& Nelson}]{prunier2025}
Prunier, M., {Hlavacek-Larrondo}, J., Pillepich, A., Lehle, K., \& Nelson, D. 2025, \href{http://dx.doi.org/10.1093/mnras/stae2743}{\color{blue}MNRAS}, 536, 3200

\bibitem[{Rasia {et~al.}(2015)Rasia, Borgani, Murante, Planelles, Beck, Biffi, {Ragone-Figueroa}, Granato, Steinborn, \& Dolag}]{rasia2015}
Rasia, E., Borgani, S., Murante, G., {et~al.} 2015, \href{http://dx.doi.org/10.1088/2041-8205/813/1/L17}{\color{blue}ApJ}, 813, L17

\bibitem[{{Rodriguez-Gomez} {et~al.}(2015){Rodriguez-Gomez}, Genel, Vogelsberger, Sijacki, Pillepich, Sales, Torrey, Snyder, Nelson, Springel, Ma, \& Hernquist}]{rodriguez-gomez2015}
{Rodriguez-Gomez}, V., Genel, S., Vogelsberger, M., {et~al.} 2015, \href{http://dx.doi.org/10.1093/mnras/stv264}{\color{blue}MNRAS}, 449, 49

\bibitem[{Rohr {et~al.}(2024{\natexlab{a}})Rohr, Pillepich, Nelson, Ayromlou, P{\'e}roux, \& Zinger}]{rohr2024a}
Rohr, E., Pillepich, A., Nelson, D., {et~al.} 2024{\natexlab{a}}, The Cooler Past of the Intracluster Medium in {{TNG-Cluster}}

\bibitem[{Rohr {et~al.}(2024{\natexlab{b}})Rohr, Pillepich, Nelson, Ayromlou, \& Zinger}]{rohr2024}
Rohr, E., Pillepich, A., Nelson, D., Ayromlou, M., \& Zinger, E. 2024{\natexlab{b}}, \href{http://dx.doi.org/10.1051/0004-6361/202348583}{\color{blue}A\&A}, 686, A86

\bibitem[{Rossetti \& Molendi(2010)}]{rossetti2010}
Rossetti, M. \& Molendi, S. 2010, \href{http://dx.doi.org/10.1051/0004-6361/200913156}{\color{blue}A\&A}, 510, A83

\bibitem[{Ruppin {et~al.}(2021)Ruppin, McDonald, Bleem, Allen, Benson, Calzadilla, Khullar, \& Floyd}]{ruppin2021}
Ruppin, F., McDonald, M., Bleem, L.~E., {et~al.} 2021, \href{http://dx.doi.org/10.3847/1538-4357/ac0bba}{\color{blue}ApJ}, 918, 43

\bibitem[{Sanchez {et~al.}(2021)Sanchez, Tremmel, Werk, Pontzen, Christensen, Quinn, Loebman, \& Cruz}]{sanchez2021}
Sanchez, N.~N., Tremmel, M., Werk, J.~K., {et~al.} 2021, \href{http://dx.doi.org/10.3847/1538-4357/abeb15}{\color{blue}ApJ}, 911, 116

\bibitem[{Sanders {et~al.}(2018)Sanders, Fabian, Russell, \& Walker}]{sanders2018}
Sanders, J.~S., Fabian, A.~C., Russell, H.~R., \& Walker, S.~A. 2018, \href{http://dx.doi.org/10.1093/mnras/stx2796}{\color{blue}MNRAS}, 474, 1065

\bibitem[{Sanderson {et~al.}(2006)Sanderson, Ponman, \& O'Sullivan}]{sanderson2006}
Sanderson, A. J.~R., Ponman, T.~J., \& O'Sullivan, E. 2006, \href{http://dx.doi.org/10.1111/j.1365-2966.2006.10956.x}{\color{blue}MNRAS}, 372, 1496

\bibitem[{Springel(2010)}]{springel2010}
Springel, V. 2010, \href{http://dx.doi.org/10.1111/j.1365-2966.2009.15715.x}{\color{blue}MNRAS}, 401, 791

\bibitem[{Springel {et~al.}(2018)Springel, Pakmor, Pillepich, Weinberger, Nelson, Hernquist, Vogelsberger, Genel, Torrey, Marinacci, \& Naiman}]{springel2018}
Springel, V., Pakmor, R., Pillepich, A., {et~al.} 2018, \href{http://dx.doi.org/10.1093/mnras/stx3304}{\color{blue}MNRAS}, 475, 676

\bibitem[{Springel {et~al.}(2001)Springel, White, Tormen, \& Kauffmann}]{springel2001}
Springel, V., White, S. D.~M., Tormen, G., \& Kauffmann, G. 2001, \href{http://dx.doi.org/10.1046/j.1365-8711.2001.04912.x}{\color{blue}MNRAS}, 328, 726

\bibitem[{Terrazas {et~al.}(2020)Terrazas, Bell, Pillepich, Nelson, Somerville, Genel, Weinberger, Habouzit, Li, Hernquist, \& Vogelsberger}]{terrazas2020}
Terrazas, B.~A., Bell, E.~F., Pillepich, A., {et~al.} 2020, \href{http://dx.doi.org/10.1093/mnras/staa374}{\color{blue}MNRAS}, 493, 1888

\bibitem[{Truong {et~al.}(2024)Truong, Pillepich, Nelson, Zhuravleva, Lee, Ayromlou, \& Lehle}]{truong2024}
Truong, N., Pillepich, A., Nelson, D., {et~al.} 2024, \href{http://dx.doi.org/10.1051/0004-6361/202348562}{\color{blue}A\&A}, 686, A200

\bibitem[{Valdarnini \& Sarazin(2021)}]{valdarnini2021}
Valdarnini, R. \& Sarazin, C.~L. 2021, \href{http://dx.doi.org/10.1093/mnras/stab1126}{\color{blue}MNRAS}, 504, 5409

\bibitem[{Vikhlinin {et~al.}(2006)Vikhlinin, Kravtsov, Forman, Jones, Markevitch, Murray, \& Van~Speybroeck}]{vikhlinin2006}
Vikhlinin, A., Kravtsov, A., Forman, W., {et~al.} 2006, \href{http://dx.doi.org/10.1086/500288}{\color{blue}ApJ}, 640, 691

\bibitem[{Villalba {et~al.}(2024)Villalba, Dolag, \& Biffi}]{villalba2024}
Villalba, J. A.~G., Dolag, K., \& Biffi, V. 2024, How the Cool-Core Population Transitions from Galaxy Groups to Massive Clusters: {{A}} Comparison of the Largest {{Magneticum}} Simulation with {{eROSITA}}, {{XMM-Newton}}, {{Chandra}} and {{LOFAR}} Observations

\bibitem[{Weinberger {et~al.}(2017)Weinberger, Springel, Hernquist, Pillepich, Marinacci, Pakmor, Nelson, Genel, Vogelsberger, Naiman, \& Torrey}]{weinberger2017}
Weinberger, R., Springel, V., Hernquist, L., {et~al.} 2017, \href{http://dx.doi.org/10.1093/mnras/stw2944}{\color{blue}MNRAS}, 465, 3291

\bibitem[{Weinberger {et~al.}(2018)Weinberger, Springel, Pakmor, Nelson, Genel, Pillepich, Vogelsberger, Marinacci, Naiman, Torrey, \& Hernquist}]{weinberger2018}
Weinberger, R., Springel, V., Pakmor, R., {et~al.} 2018, \href{http://dx.doi.org/10.1093/mnras/sty1733}{\color{blue}MNRAS}, 479, 4056

\bibitem[{Zinger {et~al.}(2020)Zinger, Pillepich, Nelson, Weinberger, Pakmor, Springel, Hernquist, Marinacci, \& Vogelsberger}]{zinger2020}
Zinger, E., Pillepich, A., Nelson, D., {et~al.} 2020, \href{http://dx.doi.org/10.1093/mnras/staa2607}{\color{blue}MNRAS}, 499, 768

\end{thebibliography}


\begin{appendix}

\section{Identifying transformations} \label{appendix_deriv}

\begin{figure}[h]
    \centering
	\includegraphics[width=0.48\textwidth]{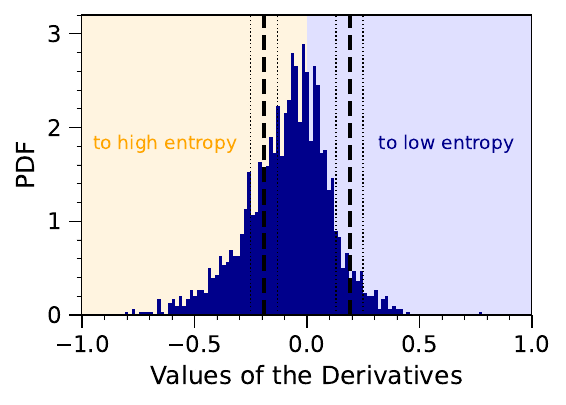}
     \caption{Histogram of all local extrema identified in the smoothed derivative of the time evolution of the central entropy for all halos in TNG-Cluster (see Fig.~\ref{fig:K0vsTime} for an exemplary illustration). Negative (positive) derivatives correspond to transformations to higher (lower) core entropy. We find 1874 of such extrema across the whole halo population. The distribution has a mean of -0.08 and a standard deviation of 0.18. For our analysis we consider only significant extrema above a value of 0.19, indicted by the vertical dashed black line. Thus, we select 478 negative and 97 positive derivatives/transformations for our analysis. The thinner dotted lines indicate variations of this threshold with values of 0.13 and 0.25.}  
    \label{fig:HistDerivatives}
\end{figure}

Our algorithm to identify cluster core transformations (Sec.~\ref{sec_findTrafos}) depends on thresholding the value of the $K_0$ time derivative in order to identify significant events. To better assess this choice, Fig.~\ref{fig:HistDerivatives} shows the histogram of all local extrema identified in the histories of all 352 clusters. The distribution peaks at -0.08, with a standard deviation of 0.19. We mark the threshold on the derivative for our fiducial analysis by the dashed line. This threshold selects 478 (26\%) negative and 97 (5\%) positive changes\footnote{The lower threshold (0.13) selects 671 (36\%) negative and 168 (9\%) positive changes; the higher one (0.35) selects 323 (17\%) negative and 51 (3\%) positive changes.}. 

\begin{figure}
    \centering
	\includegraphics[width=0.48\textwidth]{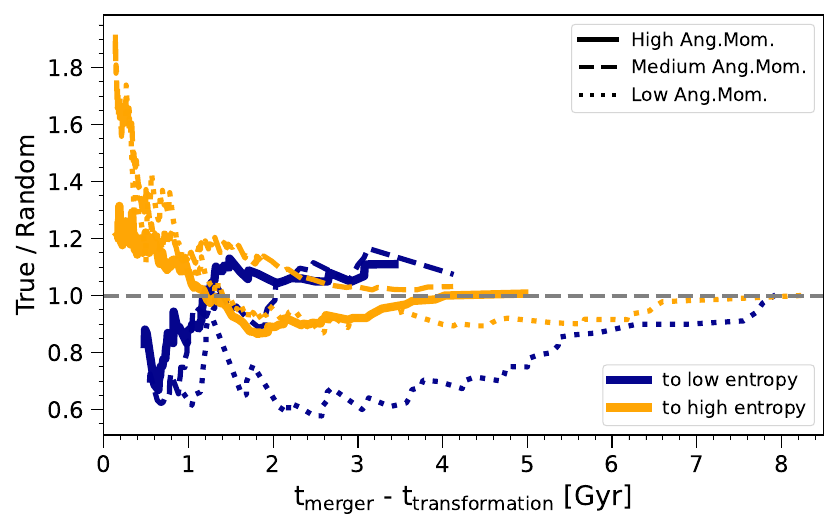}
    \caption{As in Fig.~\ref{fig:cumHistdtMerger}, considering the cumulative sum of the time difference between each transformation and its closest merger. Here we show the ratio of the true to the random result, where deviation from the unity line indicates that this sub-population has a stronger correlation between mergers and transformations than would be expected from the random shuffled control sample. While the main text studied the dependence of this trend on merger mass ratio, here we consider the dependence on the merger angular momentum.}
    \label{fig:ap_cumhist_3vt}
\end{figure}

Fig.~\ref{fig:ap_cumhist_3vt} repeats the analysis of Fig.~\ref{fig:cumHistdtMerger} and quantifies the significance of the correlation between mergers and transformations, with respect to an otherwise randomly shuffled control sample. Here we consider the dependence on the angular momentum (AM) of the merger: high (solid), medium (dashed), and low (dotted). For transformations towards higher core entropy (orange lines), the correlation within short timescales is strongest for low AM mergers.

\begin{figure}
    \centering
    \includegraphics[width=0.48\textwidth]{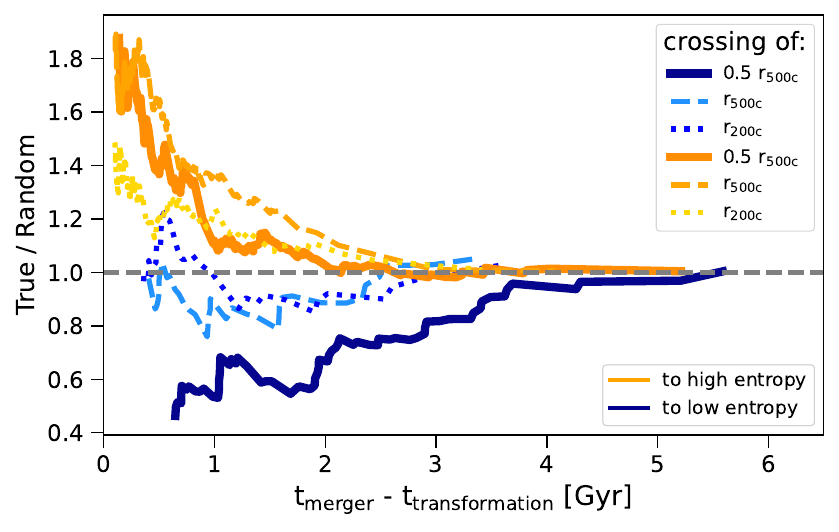}
	\includegraphics[width=0.48\textwidth]{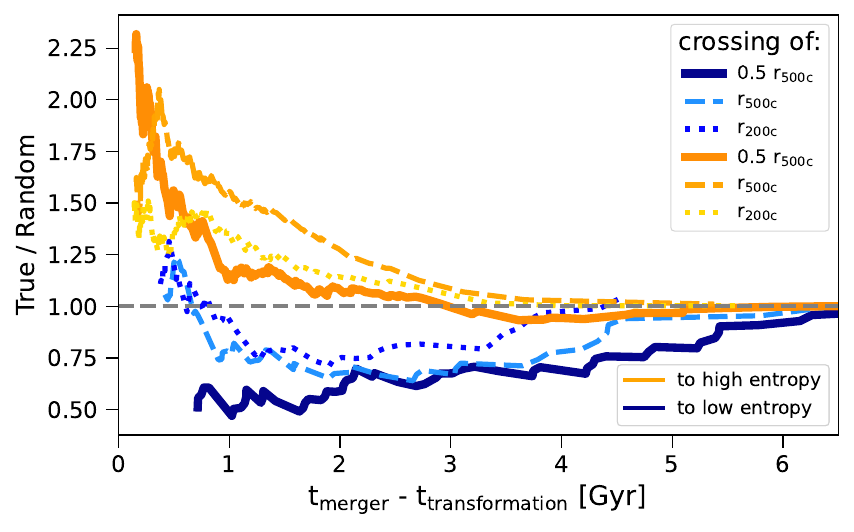}
    \caption{As in Fig.~\ref{fig:cumHistdtMerger}, but evaluating the dependence of our results on the definitions of merger mass ratio and merger time. We consider the the mass ratio defined at the time of the maximum past mass of the secondary (fiducial; top panel), versus at the time of $0.5 r_{\rm 500c}$ crossing (bottom panel). We also contrast definitions of the actual merger time: when the secondary (incoming) halo crosses $0.5 r_{\rm 500c}$ of the primary (fiducial; solid lines), versus the crossing of $r_{\rm 500c}$ (dashed lines) or $r_{\rm 200c}$ (dotted lines).}
    \label{fig:ap_cumhist_variedMRrCross}
\end{figure}

Fig.~\ref{fig:ap_cumhist_variedMRrCross} also repeats the analysis of Fig.~\ref{fig:cumHistdtMerger} in order to quantify the dependence of our results on our choice of merger time and our definition of merger mass ratio. Our fiducial merger time definition is the time that the secondary (incoming) halo crosses $0.5 r_{\rm 500c}$ of the primary (solid lines). We compare this to the $r_{\rm 500c}$ crossing time (dashed) and $r_{\rm 200c}$ crossing time (dotted). The top panel retains our fiducial definition of merger mass ratio, measured at the time of the maximum past mass of the secondary, while the bottom panel instead measures the merger mass ratio at the time of $0.5 r_{\rm 500c}$ crossing. Overall, our qualitative conclusions always hold, although their apparent statistical significance depends somewhat on details of the analysis methodology.

\section{Scenarios of Cluster Evolution} \label{sec_appendix}

\begin{figure*}
    \centering
    \includegraphics[width=0.48\textwidth]{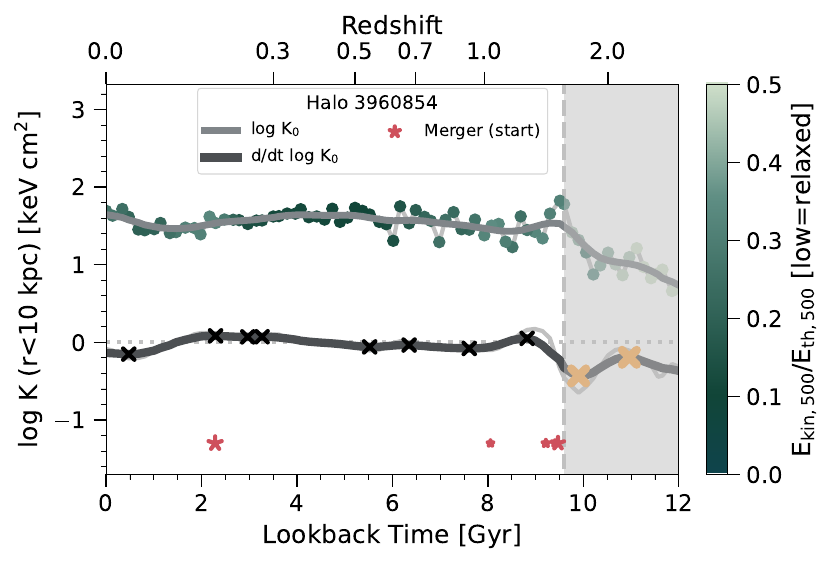}
    \includegraphics[width=0.48\textwidth]{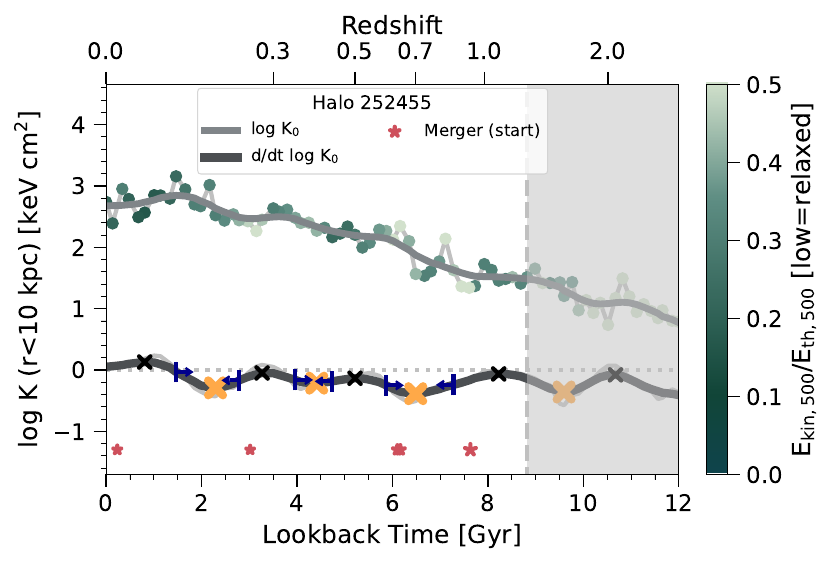}
    \includegraphics[width=0.48\textwidth]{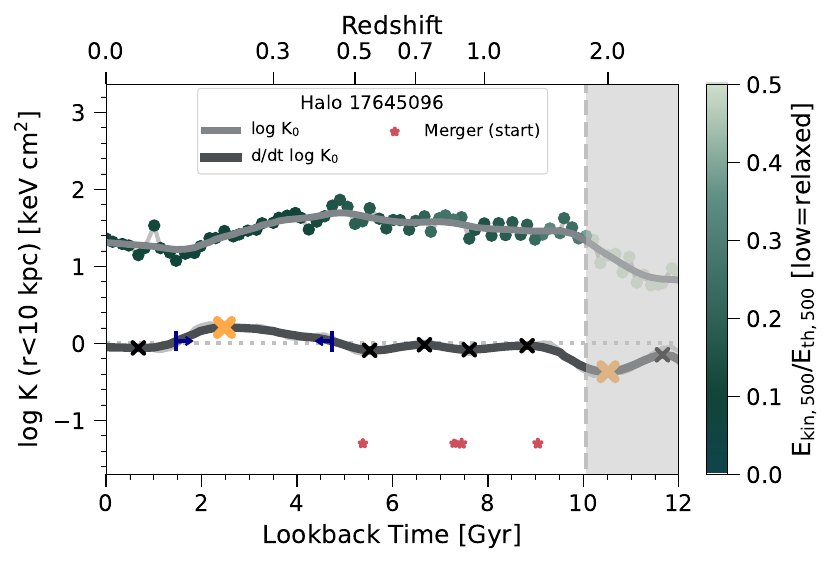}
    \includegraphics[width=0.48\textwidth]{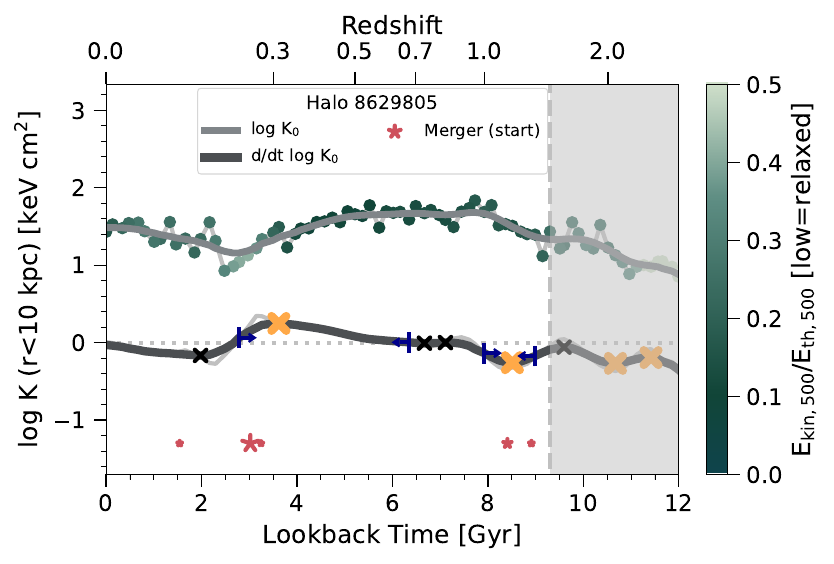}
    \includegraphics[width=0.48\textwidth]{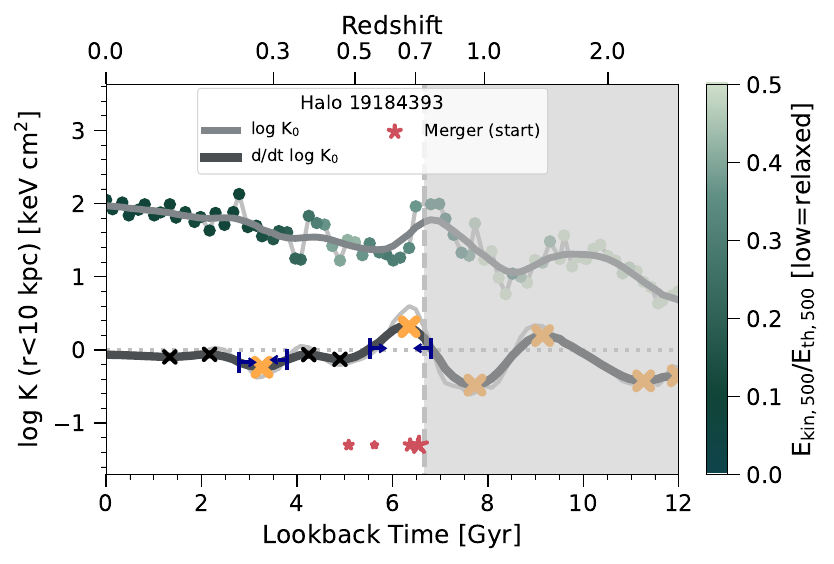}
    \includegraphics[width=0.48\textwidth]{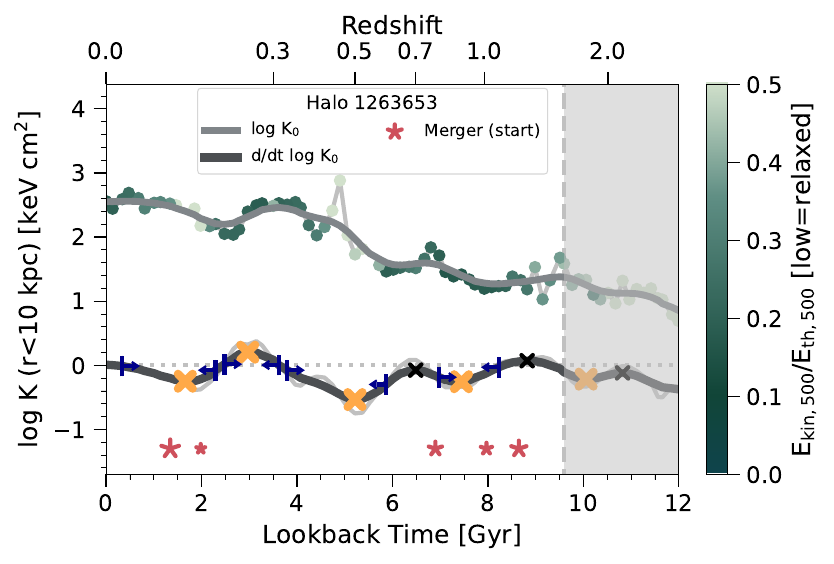}
    \caption{We identified six different archetypes in the evolution of the cluster core across the whole population in TNG-Cluster. upper left: no changes; upper right: only changes toward higher core entropy; center left: only changes toward lower core entropy; center right: temporary changes (to high then to low entropy); lower left: temporary changes (to low then to high entropy); lower right: irregular.}
    \label{fig:scenarios}
\end{figure*}

Fig.~\ref{fig:scenarios} shows six examples of central entropy evolution as a function of lookback time. Each panel is a representative example from one of our six classes (Table \ref{tab:archetypes}). Each has identical structure to Fig.~\ref{fig:K0vsTime}, in particular: the green markers show $K_0$, with color indicating relaxedness, in terms of the ratio of kinetic to thermal energy within $\rvir$. The light gray line shows the smoothed time evolution of $K_0$, while the dark gray curve shows its time derivative. Local extrema of the derivative are indicated by crosses, with larger orange crosses marking significant core transformation events. The blue markers indicate the timescales or duration of these transformations. The pink stars mark the starting points of merger events, and their size reflects the mass ratio. The formation time of each cluster is indicated by the vertical gray dashed line.

The six examples show the following behavior: no significant changes (upper left), only changes towards higher core entropy i.e. towards NCC states (upper right), only changes towards lower core entropy i.e. towards CC states (center left), temporary transformations towards higher entropy i.e. a short-lived NCC phase followed by a rejuvenated CC (center right), temporary transformations towards lower entropy i.e. a transitory CC phase followed by disruption of the CC (lower left), and irregular or complex (lower right).

\section{Low-mass cluster core transformations: TNG50}
\label{sec_tng50}

The TNG50 simulation \citep{nelson2019a,pillepich2019} is a much smaller volume than either TNG300 or TNG-Cluster. However, it contains two $\simeq 10^{14}\msun$ halos at $z=0$. The most massive undergoes a significant transformation from CC to NCC state at $z \sim 0.3$, and is also contained in a high time-resolution subbox. This offers us the opportunity to study the process of core transformation in the low-mass regime, with the additional benefit of more than 2 dex better mass resolution than TNG-Cluster.

\begin{figure*}
    \centering
    \includegraphics[width=0.48\textwidth]{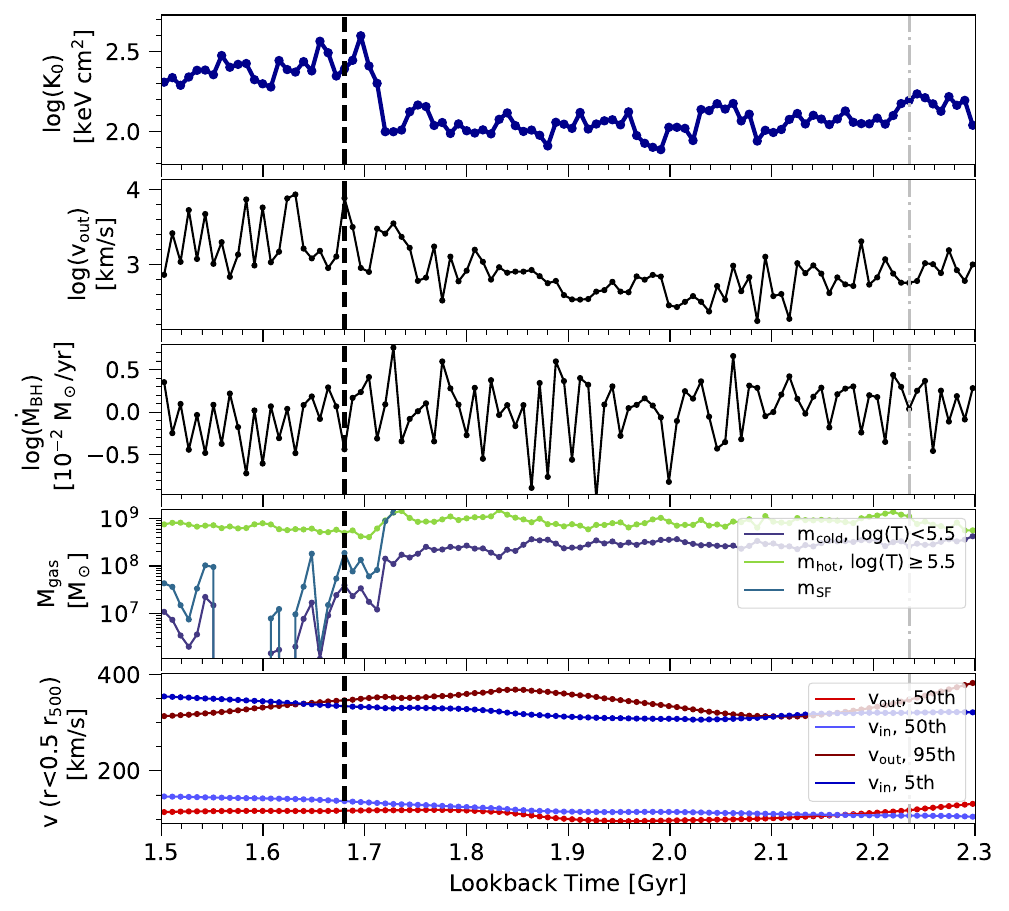}
    \includegraphics[width=0.48\textwidth]{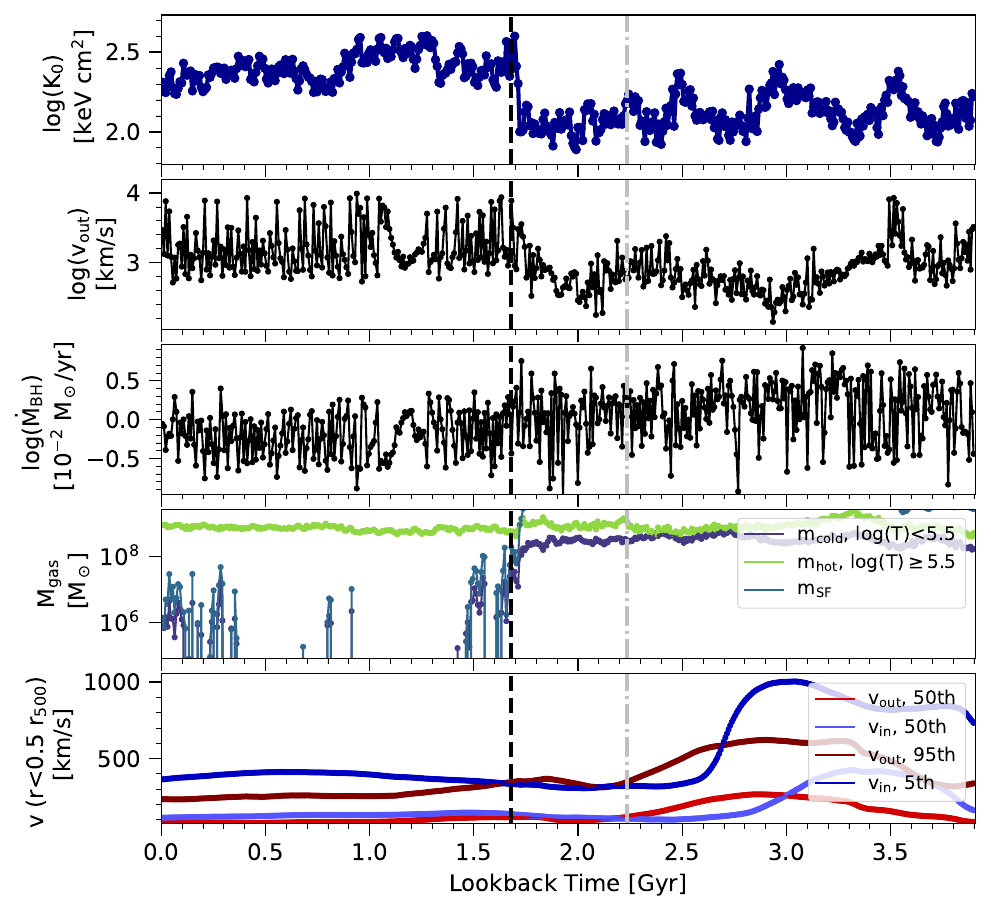}
    \caption{Time evolution of the central gas and AGN properties of a single cluster, as in Figs.~\ref{fig:timeseries} and \ref{fig:timeseries2}. Here, we perform the same analysis on the most massive halo from the TNG50 simulation. This is a low-mass cluster with $M_{\rm halo} \simeq 10^{14.1}\msun$ at $z=0$, and is also contained in a high time resolution subbox.}
    \label{fig:timeseries50}

\end{figure*}

\begin{figure*}
    \centering
    \includegraphics[width=0.48\textwidth]{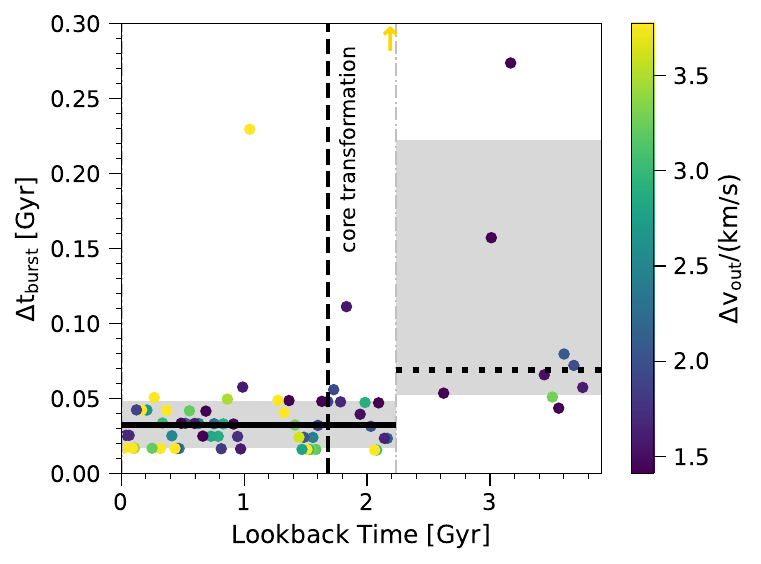}
    \caption{As in Fig.~\ref{fig:TimeDiffAGNBursts}, except here we apply an analogous analysis to the TNG50 cluster. The time between AGN feedback episodes decreases dramatically during the core transformation in comparison to the several Gyr prior.}
    \label{fig:dutycycle50}
\end{figure*}

In Fig.~\ref{fig:timeseries50} we study the time evolution of the single most massive halo in TNG50. We repeat the same analysis exactly as in Figs.~\ref{fig:timeseries} and \ref{fig:timeseries2}. Namely, the right panel shows a long time-scale view, and the left panel zooms in to the moment of the single transformation, from a CC to NCC state, that occurs at a lookback time of $\sim 1.7$\,Gyr. A merger occurs shortly before this transformation, and after the transformation is complete the cold gas reservoir is largely destroyed (fourth panel).

Fig.~\ref{fig:dutycycle50} repeats the AGN feedback duty-cycle analysis of Fig.~\ref{fig:TimeDiffAGNBursts}, but for the case of this TNG50 cluster. During the transformation the AGN activity is increased, while prior to this time the AGN activity is significantly lower. After the transformation the cold gas reservoir is destroyed.

\end{appendix}

\end{document}